\documentclass[pra, twocolumn, superscriptaddress,showpacs]{revtex4}
\usepackage{amsmath}%\usepackage{graphics}
\usepackage{graphicx}
\usepackage{dcolumn}% Align table columns on decimal point
\usepackage{bm}% bold math
\usepackage{amssymb}
\usepackage{epsfig}

\def \beq {\begin{equation}}
\def \eeq {\end{equation}}
\def \ba {\begin{eqnarray}}
\def \ea {\end{eqnarray}}
\newcommand{\ra}{\rangle}
\newcommand{\upp}{\hspace{-0.2 pt}\uparrow}
\newcommand{\downn}{\hspace{-0.2 pt}\downarrow}
\newcommand{\SV}{\hat{\vec{S}}}
\newcommand{\Sz}{\hat{S}_z}

\newcommand{\Sp}{\hat{S}_+}
\newcommand{\Sm}{\hat{S}_-}
\newcommand{\Am}{\hat{A}_-}
\newcommand{\Ap}{\hat{A}_+}

\newcommand{\Iz}{\hat{I}_z}
\newcommand{\AV}{\hat{\vec{A}}}
\newcommand{\IV}{\hat{\vec{I}}}

\def\ket#1{\left| #1\right>}
\def\bra#1{\left< #1\right|}

\def\lsim{\mathrel{\rlap{\lower4pt\hbox{\hskip1pt$\sim$}}
    \raise1pt\hbox{$<$}}}                % less than or approx. symbol
\def\gsim{\mathrel{\rlap{\lower4pt\hbox{\hskip1pt$\sim$}}
    \raise1pt\hbox{$>$}}}                % greater than or approx. symbol

\def \beq {\begin{equation}}
\def \eeq {\end{equation}}
\def \ba {\begin{eqnarray}}
\def \ea {\end{eqnarray}}

\def\lsim{\mathrel{\rlap{\lower4pt\hbox{\hskip1pt$\sim$}}
    \raise1pt\hbox{$<$}}}                % less than or approx. symbol
\def\gsim{\mathrel{\rlap{\lower4pt\hbox{\hskip1pt$\sim$}}
    \raise1pt\hbox{$>$}}}                % greater than or approx. symbol

\newcommand{\la}{\langle}
\newcommand{\hc}{\hat{c}}

\newcommand{\hd}{\hat{d}}

\begin{document}% You should use BibTeX and revtex.bst for references\bibliographystyle{apsrev}% Use the \preprint command to place your local institutional report% number on the title page in preprint mode.% Multiple \preprint commands are allowed.%\preprint{}%

\title{Fault-tolerant quantum repeaters with minimal physical resources, and implementations based on single photon emitters}  

\author{L. Childress}
\affiliation{Department of Physics, Harvard University, Cambridge,Massachusetts, 02138}

\author{J. M. Taylor}
\affiliation{Department of Physics, Harvard University, Cambridge,Massachusetts, 02138}

% moved himself here
\author{A. S. S\o rensen}
\affiliation{Department of Physics, Harvard University, Cambridge,Massachusetts, 02138}
\affiliation{ITAMP, Harvard-Smithsonian Center for Astrophysics, Cambridge,Massachusetts, 02138}
\affiliation{The Niels Bohr Institute, University of
Copenhagen, DK-2100 Copenhagen \O, Denmark}
%\affiliation{$^\ast$To whom correspondence should be addressed; E-mail: anders.sorensen@nbi.dk.}

\author{M. D. Lukin}
\affiliation{Department of Physics, Harvard University, Cambridge,Massachusetts, 02138}
\affiliation{ITAMP, Harvard-Smithsonian Center for Astrophysics, Cambridge,Massachusetts, 02138}

\date{\today}\begin{abstract}%Current techniques for quantum communication are fundamentally limited in
%range by photon attenuation in optical fibers.   
We analyze a novel method that uses fixed, minimal physical resources to achieve  generation and nested  purification of quantum entanglement for quantum communication over arbitrarily long distances, and discuss its implementation using realistic photon emitters and
photonic channels.   
In this method, we use single photon emitters with  two internal degrees of freedom formed by an electron spin and a nuclear spin to build intermediate nodes in a quantum channel. State-selective fluorescence is used for probabilistic entanglement generation between electron spins in adjacent nodes. We analyze in detail several approaches which are applicable to realistic, homogeneously broadened single photon emitters.    Furthermore, the coupled electron and nuclear spins can  be used to efficiently implement entanglement swapping and purification.   We show that these  techniques can be combined to generate high-fidelity entanglement over arbitrarily long distances.   We present a  specific protocol that functions in polynomial time and tolerates percent-level errors in entanglement fidelity and local operations.  The scheme has the lowest requirements on physical resources of any current scheme for fully fault-tolerant quantum repeaters.
 
\end{abstract}\pacs{03.67.Hk, 03.67.Mn, 78.67.Hc}\maketitle

\section{Introduction}

Quantum communication holds promise for transmitting secure
messages via quantum cryptography, and for distributing quantum
information~\cite{gisin02}.   However,  exponential 
attenuation in optical fibers
fundamentally limits the range of direct quantum communication
techniques~\cite{brassard00}, and extending them to  long distances
remains a conceptual and technological challenge.    

In principle, the limit set by 
photon losses can be overcome by
introducing intermediate quantum nodes and utilizing a so-called
quantum repeater protocol~\cite{briegel98}. 
Such a repeater creates quantum entanglement over long distances by
building a backbone of entangled pairs between closely-spaced quantum
nodes.  Performing an entanglement swap at each intermediate node
~\cite{zukowski93} leaves the outer two nodes entangled, and this
long-distance entanglement can be used to teleport quantum information
~\cite{bennett93,bouwmeester97} or transmit secret messages via quantum key
distribution~\cite{ekert91}.   Even though quantum operations are
subject to errors, by incorporating entanglement purification
~\cite{Bennett96,Deutsch96} at each step, one can extend entanglement
generation to arbitrary distances without loss of
fidelity in a time that scales polynomially with distance
~\cite{briegel98}. This should be compared to direct communication,
which scales exponentially, making it impractical for long distances.

Several approaches for physical implementation  of a quantum repeater protocol have been proposed. Early work was based on systems of several atoms trapped in high finesse optical cavities \cite{vanenk98,duan04,
blinov04}.  Such systems can form a quantum network with several quantum bits (qubits) per node, and are particularly suitable for efficient implementation of the pioneering proposal of Ref. \cite{briegel98}. 
In this approach, quantum communication over %lily removed few-
thousand kilometer distances %CHECK THIS NUMBER -- I did.  It's 1000km%
requires %approximately 
seven quantum bits per node, which must 
be coherently coupled to perform local quantum logic operations%Anders added
, i.e. a seven qubit quantum computer.  
The specific implementation of these early ideas involved the
techniques of cavity QED for interfacing stationary and photonic  
qubits and for performing the necessary quantum logic operations \cite{Bose99, Ye99}. 
Recent related work pointed out that long-distance entanglement
can be implemented via probabilistic techniques without the use of
ultra-high finesse cavities \cite{duan04,blinov04}, while local 
operations can be accomplished via short-range interactions involving
e.g. interacting trapped ions.  %Question from Anders: Shouldn't we
%add a reference here??? 
However,  few-qubit  registers 
are still technically very difficult to construct, and the difficulty increases drastically with the number of qubits involved.   At the same time, a novel approach 
based on photon  storage in atomic
ensembles~\cite{Duan01} and probabilistic entanglement is  also being actively explored.  In comparison with systems based on many-qubit nodes, this approach offers less error tolerance and requires a longer communication time. Realization
of a robust, practical system that can tolerate all expected errors
%anders added ``e'' on therefor 
remains therefore a challenging task.

In a recent paper \cite{shortpaper} we proposed a quantum repeater
protocol which could be implemented using the electronic and nuclear
degrees of freedom in single photon emitters. Here we present further
details of the proposal described in
Ref. \cite{shortpaper}, and we compare our methods to alternative
strategies.  
We show that our repeater protocol requires only two
effective quantum bits at each node. 
  This is the minimum requirement
on physical resources which still allows active error correction.    
As a specific implementation, we consider nodes formed by a single quantum emitter with two internal degrees of freedom.   
A pair of  electronic spin sublevels allows for state-selective optical excitation  (see inset in Figure~1a), and  a proximal nuclear spin provides an auxiliary memory. State-selective fluorescence is used for probabilistic entanglement generation between electronic spin degrees of freedom. We analyze in detail and compare several approaches for probabilistic entanglement generation, focussing on the feasibility of their implementation using realistic photon emitters.    Once electronic spin entanglement is generated, the coupled electron and nuclear spin at each node can  be used to efficiently implement entanglement swapping and purification.   We show that these  techniques can be combined to generate high-fidelity entanglement over arbitrarily long distances.   We present a  specific protocol that functions in polynomial time and tolerates percent-level errors in entanglement fidelity and local operations.

Our approach is stimulated by recent experimental progress in single photon generation by means of single quantum emitters, including atoms and ions as well as impurities and nanostructures in solid state devices. Although our approach is relevant
to atomic systems, such as  single atoms trapped in a
cavity~\cite{mckeever04} or single 
 trapped ions~\cite{blinov04}, it is particularly suitable for
 implementation with solid-state emitters, for example impurity color
 centers~\cite{weinfurter00,beveratos02} and quantum dots
~\cite{michler00,santori02}.  These devices offer many
 attractive features including optically accessible electronic and
 nuclear spin degrees of freedom, potential opto-electronic integrability, and
 fast operation.   
 
The paper is organized as follows.  First, we will discuss techniques for entanglement generation.  For clarity, we will present our results in the context of nitrogen-vacancy (NV) centers in diamond, and discuss alternative implementations at the end.   Realistic imperfections, such as homogeneous broadening and limited selection rules, motivate a novel entanglement generation scheme based on state-selective Rayleigh scattering and interferometry.  We calculate the success probability and entanglement fidelity for this scheme as implemented in NV centers, and compare this scheme to alternative schemes based on Raman scattering or optical $\pi$ pulses, with success conditioned on detection of one or two photons.    Next, we will show how hyperfine coupling between the electron spin and proximal nuclei permits entanglement swapping and purification.  Performing these operations in parallel and employing a nesting scheme,  we calculate the fidelity obtained and the time required to generate it as a function of distance.   In addition, we compare this scheme to pioneering proposals \cite{briegel98, Duan01} for fault-tolerant quantum repeaters.  Finally, we quantitatively discuss the feasibility of implementing a quantum repeater using NV centers, and elucidate alternative material systems which satisfy the requirements of our technique.

\section{Entanglement generation}
The initial step in our scheme is entanglement generation between the electron spins of two
emitters separated by a distance $L_0$.  
In principle, entanglement
can be generated probabilistically by a variety of means, e.g.,  Raman
scattering~\cite{cabrillo99,Bose99,browne03}
or polarization-dependent
fluorescence~\cite{blinov04}. 
%However, for the  repeater protocol  we require  storage of quantum information in the nuclear spin while we entangle the electronic spin with a different node.
% It is therefore necessary that the optical transition is
%disentangled from the nuclear spin.
However, %relevant  
%solid-state emitters
%such as color centers and doped quantum dots 
%often  do not exhibit
%appropriate selection rules,  %appropriate for Raman scattering~\cite{cabrillo99} or polarization-dependent fluorescence~\cite{blinov04}, 
%and %they typically have
%substantial homogeneous broadening on their optical transitions.
%Furthermore, 
for our repeater protocol it is essential that the
optical transition be independent of the nuclear spin
state, and 
%Specifically, as illustrated below, 
solid state emitters do not always allow Raman
scattering or polarization-dependent
fluorescence which fulfills this %the latter 
requirement. 
We therefore consider an %novel 
entanglement mechanism based on
state-selective elastic light scattering as shown in Figure~1.
Elastic light scattering places few restrictions on selection rules,
and permits nuclear-spin-independent fluorescence as we discuss below.  
% Asdiscussed for
%the specific implementations below, such elastic scattering can be
%disentangled from the nuclear spin, which is therefore undisturbed by
%the entangling operation.     

\subsection{Properties of single color centers}

\begin{figure}
  \centering
  \includegraphics[width=6cm]{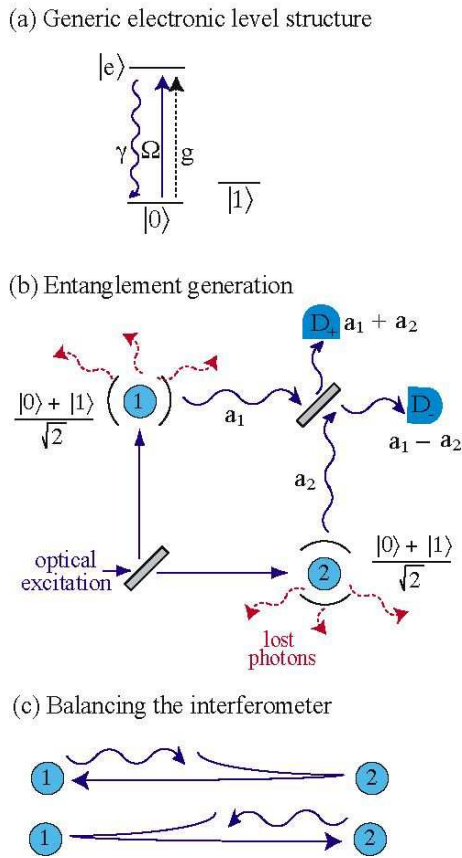}
  \caption{\it
  (a) Generic level structure showing the state-selective optical transitions and electronic spin sublevels required for entanglement generation.  (b) Setup used to create entanglement. The two emitters act as state dependent mirrors in an interferometer. The outputs of the cavities
  ($a_1$ and $a_2$) are combined on a beamsplitter. By proper
  alignment of the interferometer the photons always exit through the
  $(a_1+a_2)/\sqrt{2}$ port if both centers are in the scattering
  state $\ket{0}$. A detection of a 
  photon in the $(a_1-a_2)/\sqrt{2}$ mode thus leads to an entangled state. (c) Scheme for balancing the interferometer.  Each node is optically excited by a laser pulse which first reflects off the other node, so that the optical path lengths for the two excitation/emission paths are identical.   } 
  \label{fig:levels}
\end{figure}

\begin{figure}[htbp]\vspace{.2in}
\centerline {
\includegraphics[ width=2.5in]{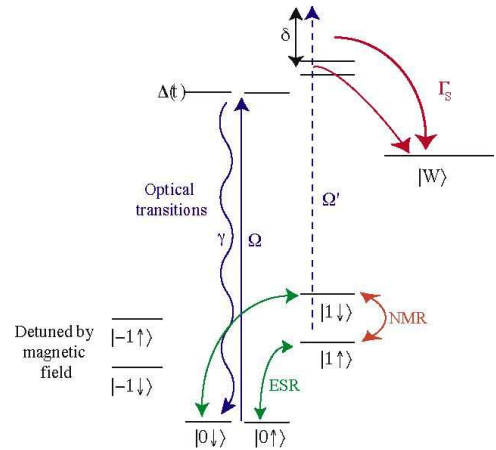}
}
\vspace{.0in}\caption{\it{The relevant electronic and nuclear states of the coupled NV center and $^{13}$C impurity nuclear spin.  The electron spin states can be coupled by ESR microwave fields near $2.88$ GHz, while the nuclear spin states can be addressed by NMR pulses on the $130$ MHz hyperfine transition.  A laser applied on resonance with the $M_s = 0$ optical transition produces strong fluorescence for the entanglement scheme; it will also weakly excite the $M_s = \pm 1$ transitions.  In our model we assume that only the $M_s = \pm 1$ states decay to the shelving state $\ket{W}$ at rate $\Gamma_S$.}}
\label{shelf}\end{figure}

Our entanglement generation scheme is applicable to a wide variety of physical systems, requiring only the simple level structure illustrated in Fig. \ref{fig:levels}a.  For clarity,  we will present it first using a concrete example: the nitrogen-vacancy (NV) center in diamond, which has the specific level structure shown in  Fig. \ref{shelf}.  This example illustrates many generic features common to other solid-state emitters.

NV centers represent a promising physical system because of their strong optical transition around 637 nm and optically accessible electron spin states.   In particular, the ground state ($A_1$ symmetry class of the $C_{3v}$ group) has three electronic spin-states which we label
$|-1\ra$, $|0\ra$ and $|1\ra$ according to their spin angular momentum
along the symmetry axis of the crystal ($M_s$).  Spin-orbit and
spin-spin effects lead to a splitting of  $|0\ra$ from  $|\pm 1\ra$ by
2.88 GHz.  Since we only require two
electronic spin states, $|0\ra$ and $|1\ra$, we isolate these two
states from the $|-1\ra$ state by either adding a small magnetic field
to shift the energy of the $|\pm 1 \ra$ state, or by using 
appropriately polarised ESR-pulses.  As spin-orbit and spin-spin effects are substantially
different for the optically excited state ($E$ symmetry class), the
strong transition from the $M_s=0$ sublevel of the ground orbital
state can be excited independently of the other $M_s$ 
states.  Although there is evidence for photo-bleaching at low
temperatures, current models indicate that  crossover into the dark
metastable state occurs primarily from the $M_s = \pm 1$ excited
states \cite{nizovtzev03}.   Furthermore, crossover into the trapping
state is a detectable error.   In the repeater protocol described below
we perform a series of measurements on the electronic spin.  During these
measurements, the dark state will not produce any
fluorescence, revealing the error.   Shelving into the metastable
state will 
thus influence the time  (see Appendix) but not
the fidelity associated with the repeater protocol.
Consequently, we 
assume that we are only near resonance with a  single state $|e\ra$
which has $M_s=0$, and neglect photo-bleaching effects.  

%Finally, it is worth mentioning that the scheme we propose relies on spatial filtering to distinguish between incident and scattered photons.  For nanometer-scale diamond nanocrystals, it may suffice to work on resonance so that the number of scattered photons is comparable to those incident.  Alternately, since NV centers break inversion symmetry it may be possible to drive a two-photon transition and filter the excitation light from the resonant scattered light by spectral means.

The electron spin degree of freedom suffices to generate entanglement between adjacent NV centers.  To propagate entanglement to longer distances, we will make use of an auxiliary nuclear degree of freedom $\{\ket{\uparrow},\ket{\downarrow}\}$ which will be used for storage of quantum information during the repeater protocol.  In NV centers, this nuclear degree of freedom can arise from a nearby
carbon-13 impurity or directly from the nitrogen-14 atom that forms
the center.  The  large energy separation between the $|0\ra$ and $|\pm 1\ra$ states
exceeds the hyperfine interaction by an order of magnitude, decoupling
the nuclear and electronic spins.  The energy levels can thus be
described by product states of the two degrees of
freedom. Furthermore, in states with $M_s=0$, the energy is
independent of the nuclear state.   Finally, a small magnetic field $\sim 10-100$ Gauss allows spectral resolution of the $M_s = \pm 1$ states without producing significant nuclear Zeeman splitting.  The optical transition between $\ket{0}$ and $\ket{e}$ is thus disentangled from the nuclear spin state.  Consequently, the nuclear spin  can be used to store entanglement while  the $|0\ra-|e\ra$ transition is used to generate another entangled pair of electron spins.  

\subsection{Entanglement protocol}
%We consider the situation in  which
To implement the entanglement scheme,
each NV center is placed inside a photonic cavity, whose output is coupled to a
single-mode photonic fiber 
(note, however, that cavities are not essential for this proposal, see below). 
Fibers from adjacent NV centers enter two ports of a beamsplitter, and
entangled electron spin pairs are obtained conditional on detection of
a photon in an outgoing arm of the beamsplitter.

Specifically, our protocol for entanglement generation relies on
scattering on the 
closed optical transition between $|0\ra$ and $|e\ra$. This scattering
does not change the state of the NV center; the centers essentially
act as mirrors reflecting the light only if they are in the state
$|0\ra$. We assume that each of the centers is initially prepared in
the state $(|0\ra+|1\ra)/\sqrt{2}$, so that the total state is  
\begin{equation}
  \label{eq:initstate}
 |\Psi_{{\rm ini}}\ra =  \frac{1}{2}{\left(
  |00\ra+|11\ra\right)}+\frac{1}{2} {\left(|01\ra+|10\ra\right)} .
\end{equation}
Since there is no light scattering from state $|1\ra$, we can exclude
the $|11\ra$ component if we detect any scattered photons. 
In state $|00\ra$, both centers act as mirrors, so that by balancing
the interferometer in Fig.\ \ref{fig:levels} (b) we can arrange for
the photons to leave through a specific port $D_+$ .  
A photon detection in the opposite port $D_-$ can thus only arise from the
$|01\ra$ and $|10\ra$ states and produces an entangled superposition of
these two states.

Balancing and stabilizing an interferometer over tens of kilometers
as required for the implementation of this protocol
represents a considerable challenge.    Using a method analogous to
the  plug-and-play system used in quantum key
distribution~\cite{muller97}, we can reduce this requirement to
stabilization of a small interferometer locally at each detector.
Suppose that we wish to generate entanglement between repeater nodes
$R_1$ and $R_2$.  Employing fast optical switches, we excite $R_1$ by
sending a pulse of light toward $R_2$, where the light is reflected
and sent past the detector to $R_1$.  Light emitted from $R_1$ follows
the same path back to the detector.  Similarly, we excite $R_2$ by
sending a pulse of light toward a fast switch at $R_1$.   The two
paths thus cover the same path length between the nodes 
and we are insensitive to fluctuations in the path lengths as long as
these fluctuations happen on a time scale which is longer than the
time it takes for the pulses to travel between the stations. 
Alternatively one could change to a protocol which relies on the
detection of two photons instead of one. In such protocols the
sensitivity to changes in the path lengths can be reduced considerably
\cite{simon,anders98,vanenk97}. 
 
\subsection{Entanglement fidelity in the presence of homogeneous broadening}
We now describe this process mathematically, calculating the fidelity
of the entangled pair produced by our protocol, as well as the
probability for it to  succeed.  Our  analysis incorporates dominant
sources of error in solid-state systems;  in particular, we account
for effects of 
homogeneous broadening on the optical transition.

Our model assumes that the NV centers are excited by a weak driving
field applied between the states $|0\ra$ and $|e\ra$ with Rabi
frequency $\Omega$;  the excited states then radiatively decays at a rate $\gamma$.
To describe the effect of homogeneous broadening
on the optical transition we assume that the energy of the excited
level fluctuates with a characteristic time which is slow compared to
the optical frequency and much shorter than the 
lifetime of the excited state. In this approximation the broadening
can be described by including a time-dependent detuning $\Delta(t)$ with
white-noise characteristics: $\langle \Delta(t)\rangle = 0, \langle
\Delta(t)\Delta(t')\rangle = \Gamma \delta(t-t')$. Below we shall be
working in the limit of weak driving $\Omega\ll\gamma+\Gamma$. In this
limit the light emitted from a center consists of two
contributions: %anders added ``(`` 
(i) a coherent part centered around the frequency of the driving laser
   with a width given by the width of the driving laser, and 
(ii) an
   incoherent part centered around the frequency of the transition
   with a frequency width of $\gamma+\Gamma$. 
%anders added ``w'' in two
The relative weight of these two contributions is $\gamma:\Gamma$. With
considerable broadening of the optical transition
$\Gamma\gtrsim\gamma$ it is therefore essential to filter out the
incoherent scattered light with a frequency filter to get a high
fidelity.  To filter out the incoherent light and obtain a high
collection efficiency we assume that the centers interact with an optical
cavity with a coupling constant $g$ and a decay rate $\kappa$. We
emphasize, however,
that good cavities are not essential for our proposal: we only require
sufficient collection efficiency and frequency selectivity, which
could also be obtained by collecting the light with a lens and
sending it through a narrow frequency filter.
% We assume
%that  a weak driving field is applied between the states $|0\ra$ and
%$|e\ra$ with Rabi frequency $\Omega$. 
In general the weak drive may be
detuned from the excited state, which would simplify the filtering of
coherent from incoherent light.  However, off-resonant excitation
would require a stronger driving field,  making it harder to avoid
stray light reaching the detectors.    
For simplicity we only discuss the situation where, on average,  the
driving field and cavity mode are resonant with the center.   
%However,
%we account for homogeneous broadening of the optical transition by
%including a time-dependent detuning $\Delta(t)$ with white-noise
%characteristics: $\langle \Delta(t)\rangle = 0, \langle
%\Delta(t)\Delta(t')\rangle = \Gamma \delta(t-t')$.    

The combined NV-center cavity system is then described by the Hamiltonian
\begin{equation}
  \label{eq:ham}
  H= \Delta(t) |e\ra\la e| +\frac{\Omega}{2}{\left( |0\ra\la
  e|+|e\ra\la 0|\right)} + g \hc^\dagger |0\ra\la
  e|+g |e\ra\la 0| \hc,
\end{equation}
where $\hc$ is the photon annihilation operator for the field in the cavity.
In the Heisenberg picture, decay terms can be included by considering
the quantum Langevin equations of motion for the atomic operators $\hat{\sigma}_{ij}  = \ket{i}\bra{j}$, 
\ba
  \label{eq:lang}
\frac{d\hc}{dt} & =& -\frac{\kappa}{2}\hc - i g \hat{\sigma}_{0e}+ {\hat F}_c \\
  \label{eq:lang2}
\frac{d\hat{\sigma}_{0e}}{ dt} &= &\left(-\frac{\gamma}{2}-i\Delta(t)\right) \hat{\sigma}_{0e} +
\nonumber \\ 
  \label{eq:lang3}
  & & i \left(g \hc +\frac{ \Omega}{2}\right)(\hat{\sigma}_{ee}-\hat{\sigma}_{00})+ {\hat F}_{0e} \\
\frac{d\hat{\sigma}_{ee}}{dt} & =& -\gamma \hat{\sigma}_{ee} + \left( i (\frac{\Omega}{2}+ g
\hc^{\dagger})\hat{\sigma}_{0e} + {\rm h.c.}\right) +  {\hat F}_{ee}
%%I CHANGED THIS EQUATION A LITTLE BIT%% %I looks ok so it is probably
%%fine, 
\ea
where the noise  ${\hat F}_c$ is the incoming vacuum noise leading to cavity decay at rate $\kappa$ and the other noise operators ${\hat F}_{0e},{\hat F}_{ee}$ represent the effect of
other optical modes that lead to decay.
%$\gamma$. (defined above)  
%%%ANDERS, CAN YOU PUT IN EXPRESSIONS FOR THE NOISE TERMS?%%%%%%%
% only in terms of other noise operators , or in terms of raising and
% lowering operators for the free fields, I don't think this will
% improve things.
%We assume that$\Delta(t)$ is slowly varying on the time scale of the optical
%frequency, allowing us to explicitly evaluate its effects. 

We obtain an appropriate solution to the quantum Langevin equations by noting that, 
in the limit of weak driving, $\Omega\ll \gamma$, there is virtually no population of the excited state,  $\hat{\sigma}_{00}-\hat{\sigma}_{ee}\approx
\hat{\sigma}_{00}$. The solution can then be written in the form
$\hc=\alpha \hat{\sigma}_{00}+$noise %% should this be $\hc=\alpha \hat{\sigma}_{0e}+$noise ??%%
and $\hat{\sigma}_{0e}=\beta \hat{\sigma}_{00}+$noise, but the
equations for $\alpha$ and $\beta$ are complicated due to the noise
$\Delta(t)$. By averaging the Langevin equations over the noise one can,
however, find simple equations for various moments of $\alpha$ and
$\beta$, and by taking steady state solutions of the averaged equations 
we find 
\begin{eqnarray}
  \label{eq:alphabeta}
  \overline{\alpha}&=&\frac{-2g\Omega}{\kappa(\gamma+\Gamma)
  (1+4g^2/\kappa(\gamma+\Gamma))}\\      
  \overline{|\alpha|^2}&=&\frac{\frac{4g^2\Omega^2} 
{\kappa^2(\gamma+\Gamma)^2}}
  {{\left (1+\frac{4g^2}{\kappa(\gamma+\Gamma)}\right)}{\left(
 1-\frac{\Gamma\kappa}{(\gamma+\Gamma)(\gamma+\kappa)}
 +\frac{4g^2}{\kappa(\gamma+\Gamma)}\right)}   
  }\\
  \overline{\beta}&=&\frac{-i\Omega} {(\gamma+\Gamma)
  (1+4g^2/\kappa(\gamma+\Gamma))}.
\end{eqnarray}
Note that in the presence of homogeneous broadening $\Gamma \neq 0$, the moments do not factor, $\overline{|\alpha|^2}\neq |\overline{\alpha}|^2$, signifying
incoherent scattering of light into the cavity. 

%In the limit of weak driving $\Omega\ll \gamma$ (below saturation) we
%may approximate $\hat{\sigma}_{00}-\hat{\sigma}_{ee}\approx
%\hat{\sigma}_{00}$, 
% and look 
%for steady state solutions of the quantum Langevin equations.  It is
%convenient to write these solutions as $\hc=\alpha \hat{\sigma}_{00}+$noise
%and $\hat{\sigma}_{0e}=\beta \hat{\sigma}_{00}+$noise, where $\alpha$
%and $\beta$ are 
%functions of the stochastic variable $\Delta(t)$ and the other system
%parameters.  By solving the equations of motions averaged over the
%noise, we can find moments of
%$\alpha$ and $\beta$, noting in particular that
%$\overline{|\alpha|^2}\neq |\overline{\alpha}|^2$. 

We now apply the entanglement generation protocol, and use our mathematical model to predict the average density matrix components of the NV center electron spins.
In our scheme, we combine the output of the two cavities on a
beamsplitter and select the desired entangled state by conditioning on
a click in detector $D_-$,
described by the photon annihilation operator $\hd_-=\sqrt{\zeta
\kappa/2}( \hc_1-\hc_2)$.  Here, subscripts one and two refer to the
two NV-centers we are trying to 
entangle, $\zeta$ is the total collection and detection
efficiency for photons leaving the cavity,
 and we have omitted the contribution from vacuum noise
operators. 
%The total detection efficiency for the light leaving the cavity
  To describe the effect of the detection, we use
the quantum jump formalism\cite{jump}.    If the system starts out in
state $\ket{\Psi_{\rm init}}$, the density matrix element $\rho_{i,j}$
at time $t$ can be  found by  
\begin{equation}
\label{eq:jump}
\rho_{i,j}(t)=\langle \Psi_{{\rm init}}| \hd_-(t)^\dagger |j\ra\la i|
 \hd_-(t)  |\Psi_{{\rm init}}\ra %Question for Anders from Lily:  I'm
% not familiar with using quantum jump formalism with density matrices,
% but comparing it to the stochastic wavefunction approach i would
% think should there be a  \delta t in the numerator? You are right, anders
\delta t / \delta P,
\end{equation}
where the time argument $t$ is included to emphasise the
time dependent Heisenberg operators, and where $\delta P$ is the
probability to have 
a click  during a time $\delta t$,  
\begin{equation}
  \label{eq:dp}
 \delta P =\langle \Psi_{{\rm init}}| \hd_-(t)^\dagger \hd_-(t)
 |\Psi_{{\rm init}}\ra \delta t.
\end{equation}

Our entanglement generation scheme relies on interference to eliminate $D_-$ detection events coming from the initial state $\ket{00}$.   However, according to our formalism, if we start out in an initial state $|00\ra$ the probability to have a
click is given by 
\begin{eqnarray}
  \label{eq:start00}
  \delta P&=& \delta t\la
  00| \overline{\hd_-^\dagger\hd_-}|00\ra\nonumber \\
&=&\kappa\zeta\delta t {\left(
\overline{|\alpha|^2}- |\overline{\alpha}|^2\right)}, 
\end{eqnarray}
where we assume the noise is independent for the two centers. This
expression vanishes only for coherent scattering of light into
the cavity, i.e.~$\overline{|\alpha|^2}=|\overline{\alpha}|^2$ or $\Gamma = 0$.   
In the presence of broadening, there is a finite probability that light will
be detected from the $|00\ra$ state.   Similarly, $\Gamma >0$ leads to a finite probability for incoherent
scattering from $|01\ra$ and $|10\ra$.   Homogeneous broadening thus reduces the fidelity
($F=\la \Psi_{{\rm ideal}}|\rho|\Psi_{{\rm ideal}}\ra$, (where
$|\Psi_{{\rm ideal}}\ra$ denotes the ideal entangled state) by 
\begin{eqnarray}
  \label{eq:finitefiltering}
  1-F&=&\frac{3}{2} {\left(1- \frac{|\overline{\alpha}|^2}
  {\overline{|\alpha|^2}} \right)}\nonumber \\
&=&\frac{3}{2} \frac{\Gamma}{\gamma+\Gamma}
\frac{\kappa}{\gamma+\kappa} \frac{1}{1+4g^2/\kappa(\gamma+\Gamma)}.
\end{eqnarray}
Here we are interested in the limit where the fidelity is close to
unity and we shall therefore assume $\overline{|\alpha|^2}\approx
|\overline{\alpha}|^2$ in the calculation of other noise sources
below. 

\subsection{Other errors}
In addition to the error caused by homogeneous broadening, there is also a
reduction in fidelity caused by multiple spontaneous emission events from the centers. 
This fidelity can
conveniently be expressed in terms of the total emission probability 
\beq 
P_{em}=\frac{t_0\Omega^2}{(\gamma+\Gamma) [1+4g^2/\kappa(\gamma+\Gamma)] }, 
\label{Pem}
\eeq where
$t_0$ is the duration of the applied laser pulse.
%To calculate the fidelity,  we first find the time evolution of various density matrix operators 
%\ba
%\hat{\sigma}_{00}(t)&=&\hat{\sigma}_{00}(t=0)\\
%\hat{\sigma}_{11}(t)&=&\hat{\sigma}_{11}(t=0)\\
%\overline{\hat{\sigma}_{10}}(t)&=&\hat{\sigma}_{10}(t=0)e^{(-i\Omega
%\overline{\beta}t/2)}, 
%\ea     
% where we have again omitted the vacuum noise.   With these
 %expressions, we can calculate 
 %the fidelity conditioned on a click in the detector, 
In the absence of homogeneous broadening, multiple excitations result in a fidelity
\begin{equation}
  \label{eq:fidelnodestinguish}
     F=\frac{1}{2}+\frac{{\rm e}^{-P_{em}(1-\epsilon/2)}}{2}
\end{equation}
and success probability
\beq
P=(1-\exp(-\epsilon P_{em}/2))/2.
\eeq
The total
collection efficiency can be expressed  as $\epsilon=\zeta P_{{\rm cav}}$
with the probability to emit into the cavity given by
\begin{equation}
  \label{eq:pcav}
  P_{{\rm cav}}=\frac{4g^2/\kappa(\gamma+\Gamma)}
  {1+4g^2/\kappa(\gamma+\Gamma)}. 
\end{equation}

This treatment has neglected the possibility of distinguishing
multiple photon detection events.    If our detector can resolve photon number, we can
use the information to improve
our protocol. 
In particular, a detection in the mode
described by  $\hd_+=\sqrt{\zeta \kappa/2}( \hc_1+\hc_2)$+noise has no 
effect on the component of Eq.\ (\ref{eq:initstate}) that we are
interested in, since $d_+(|01\ra+|10\ra) \propto (|01\ra+|10\ra)$.
Furthermore, a detection in this plus mode contains contributions from
$|00\ra$, so it yields no useful information.  On the other hand,  
detection events in the mode described by $\hd_-$  change
the sign of the superposition state, since $\hd_-(|01\ra+|10\ra)\propto
(|01\ra-|10\ra)$, and $\hd_-(|01\ra- |10\ra)\propto
(|01\ra+|10\ra)$.  Consequently, the optimal strategy is to change the
phase of the entangled state when an even number of photons is
detected.   The resulting fidelity is
\begin{equation}
  \label{eq:fidfinite}
   F=\frac{1}{2}+\frac{{\rm e}^{-P_{em}(1-\epsilon)}}{2}.
\end{equation}

Finally, we must include the effect of
two other sources of noise: dark counts and electron spin
dephasing. In the limit of small success probability $P$ the dark
count introduces an incoherent admixture of the initial state and
thus leads to a reduction in fidelity $P_{{\rm dark}}/P$ where
$P_{{\rm dark}} = \gamma_{dc} t_0$ is the dark count probability. Electron spin
dephasing makes the state decay towards a state with fidelity 1/2 at a
rate $2\gamma_e$, yielding a reduction in the fidelity of $\gamma_e$
times the total time of the experiments.   Typically, this total time will be dominated by the classical communication time between nodes, $t_c$.
%When combining all these contributions we get the expression for the reduction of the fidelity given in Eq.\ (1) in the main paper. 

Putting these considerations together, we find that the entanglement scheme succeeds with probability
$P = (1/2)\left(1-e^{-P_{\rm{em}}\epsilon/2}\right)\approx \epsilon P_{\rm{em}}/4$,
producing the state $|\Psi_-\rangle$ in time $T_0 \approx (t_0+t_c) /P$ with fidelity
\ba
F_0 &=& \frac{1}{2}\left(1+e^{-P_{\rm{em}}\left(1-\epsilon\right)}\right)
-  \gamma_e(t_0 +  t_c)\nonumber \\ && -  \gamma_{dc}\frac{t_0}{P} -
\frac{3}{2}\frac{\Gamma}{\Gamma + \gamma}\frac{\kappa}{\kappa +
\gamma}\frac{1}{1+4g^2/\kappa(\gamma + \Gamma)}. 
\label{fidelity}
\ea
% Here, the first term can be derived from Eq.~(\ref{coherentstate}); the second term accounts for electron spin dephasing (at rate $\gamma_e$) during the time required for excitation $t_0$ and classical communication $t_c$;  the third term accounts for detector dark counts at rate $\gamma_{dc}$;  the last term arises from homogeneous broadening.
 %  the last term results from dephasing
% on the optical transition which adds homogeneous broadening $\Gamma$
% to the natural linewidth $\gamma$.  This final term results from
% including homogeneous broadening of the solid state emitter by a
% detuning term $\Delta$ with white-noise characteristics
% ($\mean{\Delta(t)\Delta(t')} = \Gamma
% \delta(t-t')$)~\cite{andersXX}. 
%JMT: Anders, is there a particular reference we might cite for this, or is it just
% our upcoming PRA?
For realistic emitters  placed into a cavity with either a narrow
linewidth %much more narrow than the natural linewidth
$\kappa\ll\gamma$ or a large Purcell factor  $4g^2/(\kappa(\gamma + 
\Gamma))\gg 1$, 
%a modest cavity with vacuum Rabi
%coupling $g$ and linewidth $\kappa$ with  $4g^2/(\kappa(\gamma +
%\Gamma))>1$, 
 the first two terms should dominate the error. 
 %Both of these contributions introduce phase errors. 

\section{Comparison to other entanglement generation schemes}

The entanglement generation scheme that we have presented so far is
the scheme that we believe to be best suited to NV centers. 
For other systems, this may not be the case.   %The repeater protocol is of course independent of the exact method used to create the entanglement as long as the nuclear degrees of freedom are not disturbed during the process.
In particular, the presented scheme has two primary drawbacks: (1)
it relies on resonant scattering, making it difficult to filter fluorescence photons from the applied laser field; (2) to avoid loss of fidelity from incoherent scattering, one must detect only a
narrow frequency interval in the scattered light.    Other
entanglement methods present different problems which may prove easier
to resolve
or other methods may be better suited for different physical systems.  Consequently, we now briefly compare the resonant scattering  scheme presented above to alternate techniques.

\subsection{Raman transitions} 
 One of the first
schemes considered for probabilistic 
entanglement generation \cite{cabrillo99,Bose99,browne03} used Raman transitions in three level atoms.    In such schemes, an electron spin flip between non-degenerate ground states $\ket{0}$ and $\ket{1}$ is associated with absorption of a laser photon and emission of a frequency shifted Raman photon.  After interfering the emission from two atoms, detection of a Raman photon projects the two-atom state onto a state sharing at least one flipped spin.
%For example, suppose both atoms start out in a ground state $|1\ra$, and are illuminated by a laser which couples $|1\ra$ and $|e\ra$. Spontaneous emission from state $|e\ra$ to $|0\ra$ produces a single, frequency-shifted photon while simultaneously flipping the atomic spin.   
To avoid the possibility that both atoms emitted a Raman photon, the
emission probability must be quite small $P_{{\rm em}}\ll 1$.   In
this limit, a photon detection event in detector $D_\pm$ results in an
entangled spin state $|\Psi_\pm\ra$. 

The Raman scheme can be implemented using either a weak drive between
states $|1\ra$ and $|e\ra$ or with a short strong pulse which puts a
small fraction of the population into $|e\ra$. Since the latter is
equivalent to the single detection $\pi$-pulse scheme discussed below, 
here we consider only weak driving. The system can now be treated using the quantum Langevin-quantum jump approach formulated above.  
As before, homogeneous broadening on the optical transition leads to an incoherent contribution to the Raman scattered light, which reduces the entanglement fidelity in a manner similar to
Eq.~(\ref{eq:finitefiltering}). Again, for optimal fidelity the coherent part should
be isolated with a narrow frequency filter. 
If we assume perfect filtering and a small collection 
efficiency $\epsilon\ll 1$ the
fidelity conditioned on a click is given by 
\begin{equation}
  \label{eq:fidelityraman}
  F=1-P_{{\rm em}}
\end{equation}
with success probability $P=P_{{\rm em}}\epsilon.$% 2$???????????. 

In the limit of large fidelity $F\approx 1$, the Raman scheme has a
success probability which is a factor of 4 higher than for our interferometric scheme. Furthermore, the Raman scheme has
the advantage that stray light may  be spectrally filtered from the Raman photons.   Nevertheless, because of the hyperfine
interaction in state $|1\ra$, the transition 
frequency from $|1\ra$ to $|0\ra$ depends on the nuclear spin state.  
The associated detrimental effect on the nuclear coherence could potentially be avoided by using simultaneous transitions from
$|1\ra|\uparrow\ra$ and $|-1\ra|\downarrow\ra$, which are degenerate. 
To our knowledge, however, fluorescence between $|e\ra$ and
$|1\ra$ has not be observed, making it uncertain whether the Raman
scheme can be implemented 
for the NV-centers \endnote{Raman transitions have been observed in a
strong magnetic field by using the hyperfine interaction to mix the
$|0\ra$ and $|1\ra$ states, but since such mixing involves the nuclear
degree of freedom it is not applicable in the present context.}.
%anders moved the ``.'' 

\subsection{$\pi$ pulses}

Time-gated detection offers an alternate method for distinguishing scattered photons from stray incident light.  If an atom or NV center is excited by a sufficiently short, strong laser pulse, its population is coherently driven into the excited state.  The excited state $|e\ra$ then decays on a time scale
$1/\gamma$.  When the decay time is much longer than the incident pulse length, the excitation
light and the photon emitted from the atom are separated in time, and can thus be distinguished.   
Entanglement is then generated conditional on the detection of one or two photons, as elucidated below.

\subsubsection{Single detection} 
One particularly simple method for generating entanglement using $\pi$-pulses
begins with each atom in a state 
\begin{equation}
  \label{eq:smallphi}
  \cos{\left(\phi\right)} |1\ra+   \sin{\left(\phi\right)} |0\ra.
\end{equation}
The incident $\pi$-pulse excites the optically active state $\ket{0}$ to $\ket{e}$ with unit probability, and the spontaneously emitted photons are interfered on a beamsplitter and subsequently measured (as for the Raman scheme above).
Provided $\phi\ll 1$ we can ignore the possibility that both atoms are in
state $|0\ra$. A photon detection in $D_\pm$  excludes  the state
$|11\ra$, preparing  the system in $|\Psi_\pm\ra$.  

As with other entanglement schemes, high-fidelity entanglement generation requires filtering the incoherent scattering caused by homogeneous broadening of the optical transition.    In previous sections, we have  proposed to use a frequency filter to separate the narrow peak (in frequency) of coherent scattered light from the broad incoherent background.  In the present case,
filtering can be done in the time domain.   In the excitation process a
coherence is established between $|1\ra$ and $|e\ra$, and following
the excitation this coherence (the off-diagonal density matrix
element) decays at a rate $\Gamma/2$. By only conditioning on
photons emitted a very short time after the excitation, during which the coherence
has not had time to decay, a high quality entangled pair is
produced. 

To describe this process mathematically, we again assume that the atom is placed
inside an optical cavity.  In contrast to our previous calculations,
we assume that the cavity has a broad linewidth to ensure that
generated photons leave the system as fast as possible. In the limit
$\kappa\gg g$, $\gamma$, and $\Gamma$ we can adiabatically eliminate
the cavity by setting $d \hc/dt=0$ in Eq.~(\ref{eq:lang}) so that if we omit the noise
we obtain
\begin{equation}
  \label{eq:hcadiabat}
  \hc(t)=\frac{2ig}{\kappa}\hat{\sigma}_{0e}.
\end{equation}
Inserting this expression into Eqs.~(\ref{eq:lang2}) and (\ref{eq:lang3}) we find 
\ba
\overline{\hat{\sigma}_{0e}}(t)&=& \hat{\sigma}_{0e}(t=0) \exp{\left(-\frac
{\gamma_{{\rm eff}}+\Gamma}{2} t \right)} \\
\overline{\hat{\sigma}_{ee}}(t)&=& \hat{\sigma}_{ee}(t=0) \exp{\left(- 
 \gamma_{{\rm eff}} t\right)},
\ea 
where the effective decay rate $\gamma_{{\rm eff}}$ is the decay rate
enhanced by the Purcell effect
\begin{equation}
  \label{eq:purcell}
  \gamma_{{\rm eff}}=\gamma{\left( 1+\frac{4 g^2}
  {\kappa\gamma}\right)}. 
\end{equation}

To find the fidelity of the entangled state created with this method
we again use Eqs.~(\ref{eq:jump}) and (\ref{eq:dp}). For simplicity we
only work in the limit of small collection efficiency $\epsilon\ll
1$. Conditioned on a click at time $t$ after the excitation, the
fidelity of the entangled state is 
\begin{equation}
  \label{eq:fidelitypit}
  F=\cos^2(\phi){\left(\frac{1}{2}+\frac{1}{2}{\rm e}^{-\Gamma t}\right)}
\end{equation}
and the probability to have a click during the short time
interval %anders added 
from $t$ to $t+\delta t$ is 
\begin{equation}
  \label{eq:pt}
  \delta P=2\epsilon \gamma_{{\rm eff}} \delta t \sin^2(\phi) {\rm
  e}^{-\gamma_{{\rm eff}}t} ,
\end{equation}
where the collection efficiency $\epsilon=\zeta P_{{\rm cav}}$ is again
given by the collection efficiency for the light leaving the cavity
$\zeta$,  and the probability to emit into the cavity %anders changed
%here. Note that this is not the same as before.
is now given by %%  Eq. (\ref{eq:pcav}).
\begin{equation}
  P_{{\rm cav}}=\frac{4g^2/\kappa\gamma}
  {1+4g^2/\kappa\gamma}. 
\end{equation}

The success probability for a given fidelity now depends on the
ratio between the broadening and the effective decay rate
$\Gamma/\gamma_{{\rm eff}}$.  For $\Gamma=0$, the procedure of
initially transferring population from $|1\ra$ to $|0\ra$ and then applying a $\pi$-pulse
between $|0\ra$ and $|e\ra$ is equivalent to a Raman transition, and
Eqs.~(\ref{eq:fidelitypit},\ref{eq:pt})
indeed reproduces the same relation between success probability and
fidelity given in Eq.~(\ref{eq:fidelityraman}). In the limit of small
broadening, $\Gamma\ll\gamma_{{\rm eff}}$, the $\pi$-pulse scheme is
advantageous over the interferometric scheme presented first.  In particular, for a fixed
fidelity $F\approx 1$ the success probability is a factor of 4 higher. 

In the presence of
broadening, however,  the situation is different.
To obtain a high fidelity we should detect only photons emitted within a short time $T$ following the
excitation. The average fidelity will then depend on two parameters $\phi$
and $T$.  By optimizing these two parameters we find that for
$F\approx1$ the fidelity is
\begin{equation}
  \label{eq:fidelityonpi}
  F=1-\sqrt{\frac{\Gamma}{8\gamma_{{\rm eff}}}}\sqrt{\frac{P}{\epsilon}}.
\end{equation}
Since previous expressions (\ref{eq:fidelnodestinguish}) and
(\ref{eq:fidelityraman}) for $1-F$ depended linearly on $P$,  this represents
a much faster decrease in the fidelity.  The $\pi-$pulse scheme is thus less attractive for homogeneously broadened emitters.

%% Away from the regime $F\approx 1$ we have only been able to optimize
%% $\theta$ and $T$ numerically. In Fig.~\ref{fig:comparison} we compare
%% the result of this optimization with the results for the Raman and
%% resonant scattering. Because of the different filtering methods 
%% the expression we have derived corresponds to different physical setups
%% (small and large $\kappa$) and the schemes are expressions are
%% therefore not directly comparable. To make the comparison we have assumed
%% that the collection efficiency before filtering is the same, i.e.,
%% this corresponds to a fixed value of $4g^2/\kappa\gamma_{{\rm eff}}$,
%% where $\gamma_{{\rm eff}}$ for the resonant scattering and Raman
%% schemes is given by  $\gamma_{{\rm
%% eff}}=\gamma/(1+4g^2/\kappa(\gamma+\Gamma))$.   
%% Also this assumption is equivalent to assuming that a fixed
%% fraction of the emitted light is collected and
%% subsequently goes through a narrow frequency filter in the case of
%% resonant scattering and Raman. 

%%  In the figure we plot the fidelity as function of
%% the success probability divided by a modified collection efficiency
%% $P/\epsilon'$.
%% \begin{figure}[htbp]
%%   \centering
  
%%   \caption{Comparison of different schemes. (a) No broadening (b)
%%   \Gamma=3\gamma }}  
%%   \label{fig:comparison}
%% \end{figure}

\subsubsection{Double detection} If the collection efficiency is very
high, it may be an advantage to 
rely on the detection of two photons instead of one
\cite{anders02, barrett04,simon,anders98}.   In this scheme, both atoms are initially prepared in
$(|0\ra+|1\ra)/\sqrt{2}$ and a $\pi$-pulse is applied between $|0\ra$
and $|e\ra$. Following a detection in $D_\pm$ the populations in states
$|0\ra$ and $|1\ra$ are interchanged and another $\pi$-pulse is
applied between $|0\ra$ and $|e\ra$. Conditioned on clicks following
both $\pi$ pulses we can exclude the possibility that the atoms were
initially in the same state and we are left with $|\Psi_\pm\ra$
conditioned on appropriate detector clicks.  In the absence of homogeneous broadening, this protocol produces an entangled state with fidelity F=1 with probability $P=\epsilon^2/2$.  The double-detection scheme thus avoids the multiple photon emission errors inherent in the single-detection schemes.

With broadening of the optical transition, this is no longer
the case. For $F\approx1$, the relation between fidelity and success
probability is now given by
\begin{equation}
  \label{eq:fidelitypitwo}
  F=1-\frac{\Gamma}{\gamma_{{\rm eff}}\epsilon}\sqrt{2P}.
\end{equation}
Again the fidelity decreases more rapidly with the success probability
than for the Raman and resonant scattering scheme, making it less
useful for our purpose. 

\subsection{Summary}
The best choice of scheme depends on the specific physical
situation. The two $\pi$-pulse schemes are advantageous if the
broadening is negligible.  In particular, in the limit where we can ignore all errors except the 
photon attenuation, the double detection scheme results in the highest fidelity entangled pair. %Since it produces a perfect entangled state there is no need to do purification and the total communication time will just be limited by the time it takes to connect two neighboring nodes and the time required for classical communication. In practice there will most likely be errors in the local operations and it will be necessary to do purification even if we use the double detection scheme. 
With low collection efficiency or large
distances between emitters, the double detection will have
a very small success probability because of the $\epsilon^2$ factor,
and it may be advantageous to rely on a single detection scheme. 

The $\pi$-pulse schemes are less attractive if we are limited by homogeneous broadening
of the optical transition because the fidelity decreases rapidly with
the success probability.  Better results are obtained for the
resonant scattering or Raman schemes.   When possible, the Raman scheme offers the best solution.  The frequency-shifted Raman scattering allows frequency filtering of the incoming light;  in addition the success probability is four times higher than for the resonant scattering scheme.  But, as mentioned above, it is not always possible
to drive Raman transitions
, and it may be hard to achieve Raman
transition which are independent of the nuclear spin state. 
For this reason we believe that the resonant scattering scheme is most promising in 
the particular case of NV-centers.
%The Major drawback of this scheme is the problem of
%distinguishing photons emitted from the center from other scattered
%photons. This problem could be avoided by driving the center with a
%two photon transition. 

Finally we wish to add that the calculations we have performed here
assume a specific model for the broadening (short correlation time
for the noise). With other broadening mechanisms, e.g. slowly varying
noise,  these considerations will be different.

\section{Entanglement swapping and purification}

Using one of the procedures outlined above, electron spin entanglement can be generated
between adjacent pairs of nodes.  We now discuss a means to extend the entanglement to longer distances.

\subsection{Swapping}
\begin{figure}[htbp]\vspace{.2in}
\centerline {
\includegraphics[width=3in]{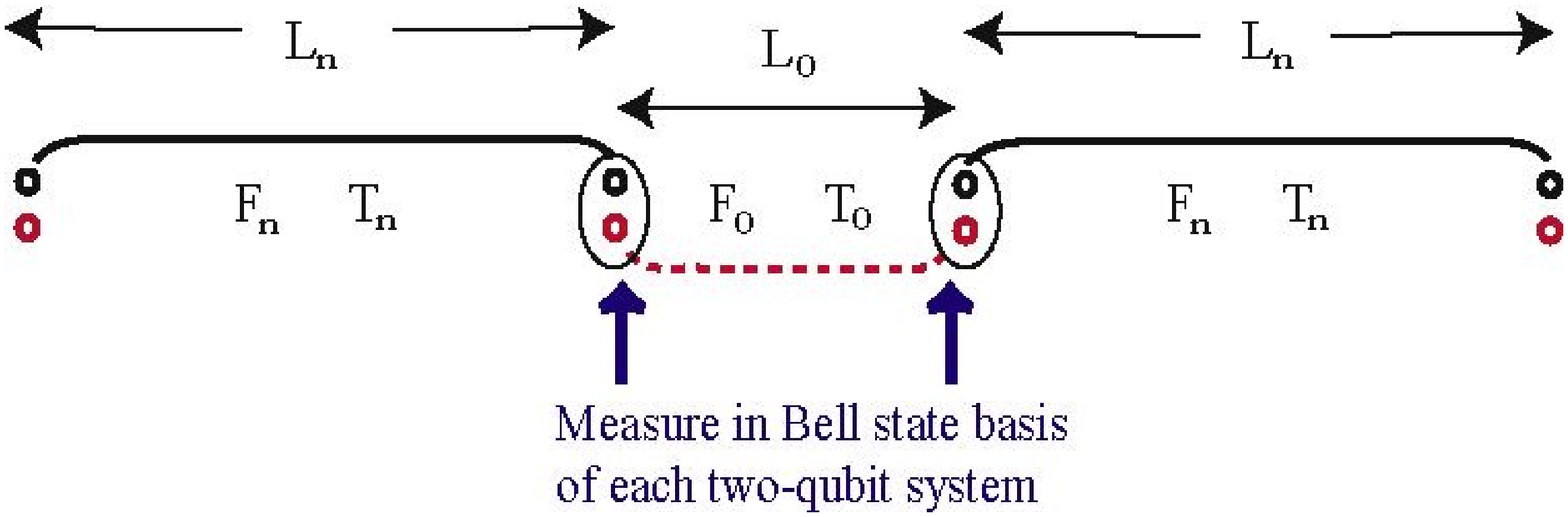}
}
\vspace{.0in}\caption{\it{Entanglement propagation by swapping. To generate an entangled nuclear spin pair (black) over the distance $L_{n+1} = 2 L_n + L_0$, we first generate nuclear entanglement over the first and second pair of repeater stations which are distance $L_n$ apart.  The electron spin (red) is then used to generate entanglement between the middle stations, separated by distance $L_0.$  
Nuclear (electron) spin entanglement is illustrated by solid (dashed) lines. Entanglement swapping is performed by measuring in the Bell state basis of the two-qubit  system.
}}
\label{scaling}\end{figure}

After entangling nearest-neighbor electron spins, the electron spin state is  mapped onto
the auxiliary nuclear spin qubit for long-term storage  
using the hyperfine interaction.  This operation leaves the electronic degree of freedom available to generate entanglement between unconnected nodes,
as illustrated in Figure~\ref{scaling}.   By combining optical detection of individual electron spin states~\cite{jelezko04} and effective two-qubit operations associated with hyperfine coupling of electronic and nuclear spins~\cite{jelezko04b}, we may projectively measure all four Bell states in the electronic/nuclear manifold associated with each emitter.  
The outcomes of the Bell state measurements reveal the appropriate
local rotations to obtain a singlet state in the remaining pair of
nuclear spins, implementing a deterministic entanglement
swap~\cite{bennett93,zukowski93}.   By performing this procedure in parallel, and iterating the process for $N
\propto \log_2{(L/L_0)}$ layers, we obtain the desired nuclear spin
entanglement over distance $L$ in a time $\propto L\log_2{(L/L_0)}$.  

\subsection{Purification}
To extend entanglement to long distances in the presence of errors,
active purification is required  at each level of the repeater scheme.    
By performing local operations and measurements, it is possible to
distill multiple entangled pairs with fidelity above some threshold
$F_{min}$ into a single entangled pair with higher
purity\cite{Bennett96, Deutsch96}.   The purification algorithm we use
is described in detail in Refs.\cite{Deutsch96, briegel98, Dur99,
DurThesis}.  For clarity we will present it in a form appropriate to the system under consideration,  %lily added
which uses repeated generation of electron
spin entangled pairs to purify a stored nuclear spin entangled pair.
Specifically, an electron spin entangled pair between stations $i$ and
$j$ is described by the density matrix diagonal components $\{a_e,
b_e, c_e, d_e\}$ in the Bell state basis $\{\ket{\Psi_-},
\ket{\Phi_+}, \ket{\Phi_-}, \ket{\Psi_+}\}$, where 
\ba
\ket{\Psi_\pm}^{(e)}_{ij} &=& \frac{1}{\sqrt{2}}\left( \ket{0_i1_j} \pm \ket{1_i0_j}\right)\\
\ket{\Phi_\pm}^{(e)}_{ij} &=& \frac{1}{\sqrt{2}}\left( \ket{0_i0_j} \pm \ket{1_i1_j}\right).
\ea
We will refer to these diagonal elements as the ``vector fidelity" $\mathcal{F}_e = \{a_e, b_e, c_e, d_e\}$, noting that the first element $(\mathcal{F}_e)_1 = a_e$ encodes the fidelity with respect to the desired singlet state.
A nuclear spin entangled pair between those stations is described by a similar vector fidelity $\mathcal{F}_n =\{a_n, b_n, c_n, d_n\}$ in the nuclear Bell basis 
\ba
\ket{\Psi_\pm}_{ij}^{(n)} &=& \frac{1}{\sqrt{2}}\left( \ket{\downarrow_i\uparrow_j} \pm \ket{\uparrow_i\downarrow_j}\right)\\
\ket{\Phi_\pm}_{ij}^{(n)} &=& \frac{1}{\sqrt{2}}\left( \ket{\downarrow_i\downarrow_j} \pm \ket{\uparrow_i\uparrow_j}\right).
\ea 
The purification protocol calls for a local rotation of each spin system at both locations:
\ba
\ket{0}_{i,j} \rightarrow \frac{1}{\sqrt{2}}\left(\ket{0}_{i,j}+i\ket{1}_{i,j}\right)\\
\ket{1}_{i,j} \rightarrow \frac{1}{\sqrt{2}}\left(\ket{1}_{i,j}+i\ket{0}_{i,j}\right)\\
\ket{\downarrow}_{i,j} \rightarrow \frac{1}{\sqrt{2}}\left(\ket{\downarrow}_{i,j}+i\ket{\uparrow}_{i,j}\right)\\
\ket{\uparrow}_{i,j} \rightarrow \frac{1}{\sqrt{2}}\left(\ket{\uparrow}_{i,j}+i\ket{\downarrow}_{i,j}\right),
\ea
followed by a two-qubit gate at each location:
\beq
\begin{array}{clcr}
\ket{\downarrow 0}_i &\rightarrow& \ket{\downarrow 0}_i\\
\ket{\downarrow 1}_i &\rightarrow& \ket{\downarrow 1}_i\\
\ket{\uparrow 0}_i &\rightarrow& -\ket{\uparrow  1}_i\\
\ket{\uparrow 1}_i &\rightarrow& -\ket{\uparrow  0}_i\\
\end{array}~~~~~
\begin{array}{clcr}
\ket{\downarrow 0}_j &\rightarrow& \ket{\downarrow 1}_j\\
\ket{\downarrow 1}_j &\rightarrow& \ket{\downarrow 0}_j\\
\ket{\uparrow 0}_j &\rightarrow& \ket{\uparrow  0}_j\\
\ket{\uparrow 1}_j &\rightarrow& \ket{\uparrow  1}_j.
\end{array}
\eeq
After these operations, the electron spin is projectively measured at both locations.  When the two electron spins are in the opposite state, the purification step succeeds, mapping the remaining nuclear spins to a diagonal state $\{a_n', b_n', c_n', d_n'\}$ with $a_n'>a_n$.   

This purification protocol can correct any type of error, but it functions best for phase errors, which correspond to diagonal elements of the form $\{f, 0, 0, 1-f\}$.   To quantify the type of error associated with our entanglement generation scheme, we define a shape parameter $\upsilon$ such that the  vector fidelity for entangled spin pairs between adjacent nodes is 
\beq
\mathcal{F}_0 = \{F_0, (1-F_0)\upsilon, (1-F_0)\upsilon, (1-F_0 )(1-2\upsilon)\}.
\label{shape}
\eeq
Note that $\upsilon \rightarrow 0$ corresponds to phase errors while
$\upsilon \rightarrow 1/3$ corresponds to a Werner state with equal
distribution of all error types. 
Note also that the assumption of diagonality imposes no restriction on
the entangled states we generate, as any off-diagonal elements in
their density matrices can be eliminated by performing random
rotations  (similar to the procedure for creating Werner states but
without the symmetrization step)\cite{DurThesis}.  Furthermore, even without a  randomization step,  the average fidelity is determined by the diagonal elements \cite{Deutsch96}.
%%%QUESTION FOR
%LILY from Anders, I remember seeing one of the purification protocols
%where it is only the diagonal which matters when we average over
%outcomes. Is that also the case here. If so isn't that what we should
%write? ANSWER:  It was mentioned in one sentence of one paper by
%Deutsch, but I don't understand the logic which led to that statement
%so I haven't mentioned it here (whereas I at least understand how one
%could randomize off-diagonal elements).  NEW QUESTION from anders:
% but in the PRL
%we give the other explanation shouldn't we be consistent?

\subsection{Errors}

In the presence of local errors in measurements and operations, the
purification and swap procedures deviate from their ideal
effect. 
To describe this we
 use the error model described in \cite{Dur99}.  Measurement errors
are quantified using a parameter $\eta$ such that measurement projects
the system into the desired state with probability $\eta$ and into the
incorrect state with probability $1-\eta$.  For example, a projective
measurement of state $\ket{0}$ would be 
\beq
P_0 = \eta\ket{0}\bra{0} + (1-\eta)\ket{1}\bra{1}.
\eeq
Errors in local operations are accounted for in a similar manner.  With some probability $p$, the correct operation is performed; otherwise one traces over the relevant degrees of freedom in the density matrix and multiplies by the identity matrix (for further details see\cite{Dur99} and references therein).   For example, the action of a two qubit operation $U_{ij}$ would become
\beq
 U_{ij}\rho U_{ij}^{\dagger} \rightarrow p ~ U_{ij}\rho U_{ij}^{\dagger} + \frac{1-p}{4} Tr_{ij}(\rho)\otimes \mathcal{I}_{ij}
\eeq
 In our calculations, we neglect errors in single qubit operations and
 focus on two-qubit errors, which are likely to yield the dominant
 contribution. 

These errors determine the level of purification which is possible
given infinitely many purification steps.  They also determine how
much the fidelity degrades during the entanglement swap procedure.  
Below we describe a repeater protocol which, compared to the original
proposal \cite{briegel98}, reduces the required
number of qubits at each repeater station at the expense of extra
connection steps.
Owing to these extra connection steps, our protocol is slightly
more sensitive to local errors than the original scheme.% proposed
%by\cite{briegel98}. 
%In essence, the reduced requirements on
%physical resources at each node exact a modest cost in increased error
%sensitivity.   

\subsection{Nesting Scheme}

Previous proposals for fault tolerant long distance quantum communication have required  larger and larger numbers of qubits at each node as the communication distance is increased.  Here we describe a nesting scheme which can be used to communicate over arbitrarily long distances while maintaining a constant requirement of only two qubits per node.  

\begin{figure}
\vspace{.2in}
\centerline {
\includegraphics[ width=3in]{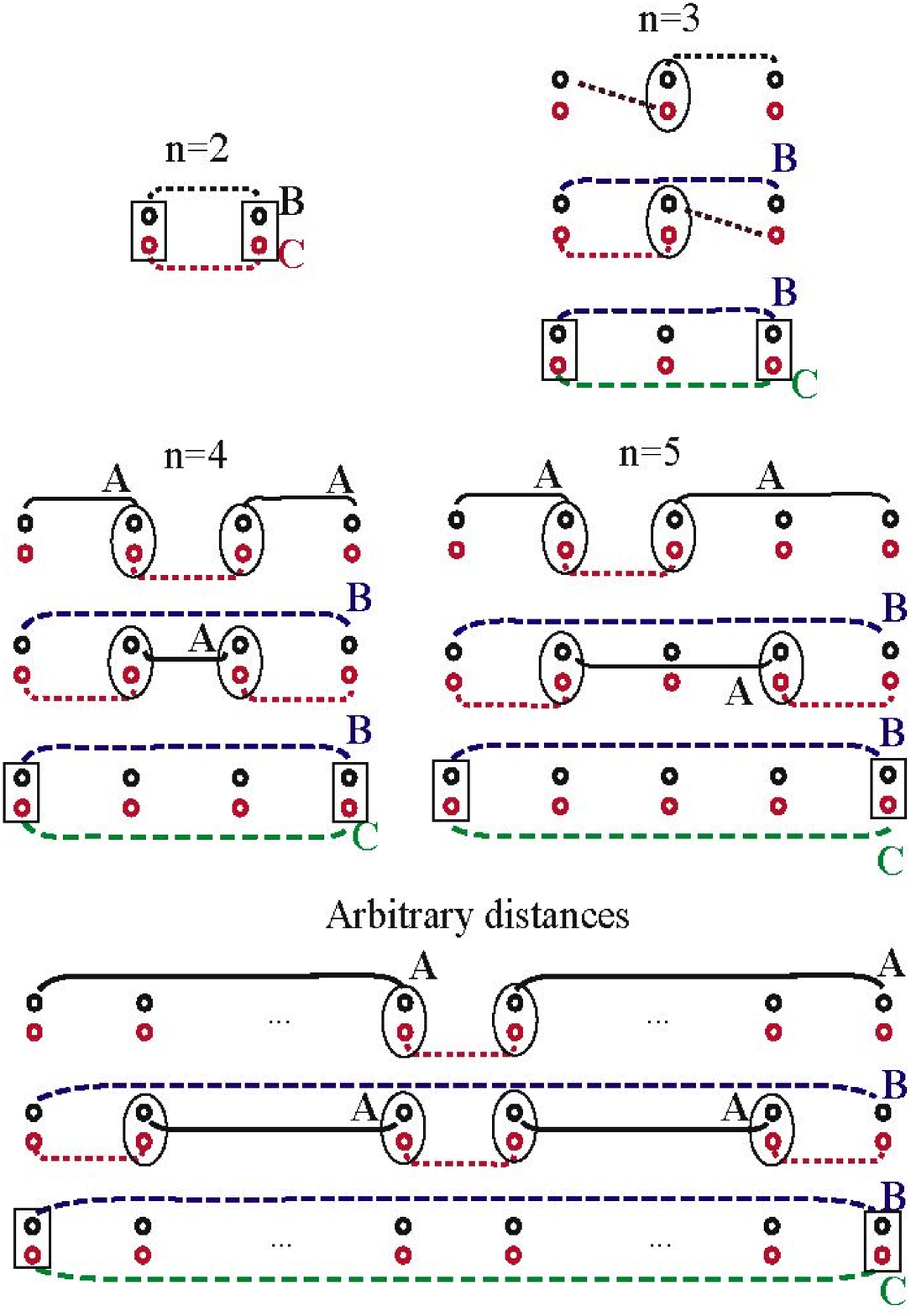}
}
\vspace{.0in}\caption{\it{Nesting scheme for generation and purification of entangled nuclear and electron spin pairs.  In each node, the nuclear spin degree of freedom is represented by the upper (black) circle, while the electron degree of freedom is represented by the lower (red) circle.  Entanglement between different nodes is represented by a line connecting them.  Ovals represent entanglement swap steps, and rectangles represent entanglement purification steps.  For $n=2$ the B and C pairs may be directly generated.  For $n\geq3$, the first step illustrates how the B pair is generated, while the remaining two steps illustrate how the C pair is generated while storing the B pair.    The arbitrary distance algorithm works for $n\geq 6$.}}
\label{first}\end{figure}

%lily moved figure up and added here
The scheme for nested entanglement purification is illustrated in Figure \ref{first}.  %lily moved this sentence
 For clarity, we will label purified pairs by``A'', pairs to be purified by ``B'', and auxiliary pairs used to perform purification by ``C". 
 Briefly, an entangled pair (``$B$") is stored in the nuclear spins while an auxiliary entangled pair (``$C$") is generated in the electron spins. 
The purification protocol described in ~\cite{Deutsch96,Dur99} is then
performed by entangling the electron and nuclear spins via the  hyperfine interaction,  
and subsequently measuring the electron
spins. Comparison of the measurement outcomes  reveals whether the
purification step was successful, resulting in a new stored pair $B$
with higher fidelity.  
After successfully repeating the procedure for $m$ purification steps, 
(a technique sometimes referred to as ``entanglement pumping" ),  the stored pair becomes a
purified (``A") pair, which can then be used to create B and C pairs over longer distances.  
We may thus generate and purify entanglement to arbitrary distances.  This procedure is analogous to the scheme in Ref.~\cite{briegel98}, but avoids the increase in the
number of qubits required for that proposal. 

%lily removed:
%Briefly, our protocol makes use of nearest-neighbor nodes to construct an electron spin entangled pair while storing a nuclear spin entangled pair.   The electronic degrees of freedom are then used to purify the nuclear spin pair.   

Mathematically, the scheme can be explained most easily using inductive arguments.   Suppose that we have a means to create and purify entanglement over $k=2, 3, \cdots, n/2$ repeater stations  ($(n+1)/2$ if $n$ is odd), and that we know the vector fidelity $\mathcal{F}_A(k)$ and the time $\mathcal{T}_A(k)$ required for each distance.  We can then determine the time required and the vector fidelity possible after purification over $n$ repeater stations.

We begin by creating two purified nuclear spin A pairs over half the distance and connecting them via a central electron spin pair of vector fidelity $\mathcal{F}_0$.  In the presence of local errors, this yields a nuclear spin B pair with vector fidelity 
\beq
\mathcal{F}_B(n) = \mathcal{C}\left(\{\mathcal{F}_A(\frac{n}{2}), \mathcal{F}_0, \mathcal{F}_A(\frac{n'}{2})\}, \eta ,p \right).
\eeq
Here,  $\mathcal{C}$ gives the vector fidelity obtained upon connecting the entangled pairs in the presence of local errors\cite{DurThesis}, and $n/2$ and $n'/2$ are understood to represent $(n-1)/2$ and $(n+1)/2$ when $n$ is odd.   The B pair is created in a time 
\beq
\mathcal{T}_B(n) = \mathcal{T}_A(\frac{n'}{2}) + T_0 +  \frac{n'}{2}t_c ,
\eeq
where $T_0$ is the time required to generate nearest-neighbour entanglement and $t_c$ is the classical communication time between adjacent stations.  We neglect the time required for local operations and measurement since these times are short compared to $T_0$ and $n t_c$.
Similarly, we can find the vector fidelity and time for the electron spin C pair 
\ba
\mathcal{F}_C(n) &=& \mathcal{C}(\{\mathcal{F}_0, \mathcal{F}_A(\frac{n}{2}-1), \mathcal{F}_0, \mathcal{F}_A(\frac{n'}{2} -1), \mathcal{F}_0\}, \eta, p)\nonumber \\
\mathcal{T}_C(n) &=& \mathcal{T}_A(\frac{n'}{2} -1) + T_0 +  (n-2)t_c .
\ea

After performing one purification step, we obtain a nuclear spin pair $A_1$, with vector fidelity determined by the purification function $\mathcal{P}$
\beq
\mathcal{F}_{A_1}(n) = \mathcal{P}(\mathcal{F}_B(n), \mathcal{F}_C(n), \eta, p).
\eeq
On average, the time required to perform this single step is
\beq
\mathcal{T}_{A_1}(n) = \frac{\left(\mathcal{T}_B(n)+\mathcal{T}_C(n)+(n-1)t_c\right)}{P_S(\mathcal{F}_B(n), \mathcal{F}_C(n))},
\eeq
where $P_S$ is the probability that the purification step succeeds.   

After $m$ successful purification steps, the vector fidelity of the nuclear spin $A_m$ pair is
\beq
\mathcal{F}_{A_m}(n) = \mathcal{P}(\mathcal{F}_{A_{m-1}}(n), \mathcal{F}_C(n), \eta, p),
\eeq
and the average time required for its creation is
\ba
\mathcal{T}_{A_m}(n) &=& \frac{\left(\mathcal{T}_{A_{m-1}}(n)+\mathcal{T}_C(n)+(n-1)t_c\right)}{P_S(m)} \\
&=& \left(\mathcal{T}_C(n) + (n-1)t_c\right)\sum_{n = 1}^m \prod_{k=n}^m\left(\frac{1}{P_S(m)}\right) 
\nonumber \\
&& +  \mathcal{T}_B(n)\prod_{k=1}^m\left(\frac{1}{P_S(m)}\right),
\ea
where $P_S(m) = P_S(\mathcal{F}_{A_{m-1}}(n), \mathcal{F}_C(n))$.
If we stop purifying at some fixed number $M$ of purification steps, then the desired vector fidelity and time over distance $n$ are given by 
\ba
\mathcal{F}_A(n) &=& \mathcal{F}_{A_M}(n) \\
\mathcal{T}_A(n) &=& \mathcal{T}_{A_M}(n). 
\ea

To complete the inductive argument, we must show that the protocol works for small distances.  There are many schemes one can use to generate and purify entanglement over shorter  distances, and one possibility is illustrated in Figure~\ref{first}.  In fact, once the physical parameters for an implementation are determined, it should be possible to optimise the few-node scheme to minimise the required time or maximise the resulting fidelity.  

\subsection{Fixed point analysis}

As the number of purification steps increases $m\rightarrow\infty$, the fidelity of the resulting entangled pair saturates.  This saturation value can be found using a fixed point analysis (as described in\cite{Dur99}) by solving for the vector fidelity $\mathcal{F}_A$ which is unchanged by further purification steps
\beq
\mathcal{F}_A = \mathcal{P}\left(\mathcal{F}_A, \mathcal{F}_C,  \eta, p \right),
\label{fp}
\eeq  
where we have explicitly included the local errors in the purification function $\mathcal{P}$.  This yields a fixed point fidelity $\mathcal{F}_{FP}(\mathcal{F}_C, \eta, p)$ which is independent of $\mathcal{F}_A$.  Since the vector fidelity $\mathcal{F_A}$ has three independent parameters characterising the diagonal elements of the density matrix, one might miss the fixed point.  However, as the number of purification steps increases our simulations do indeed approach the calculated fixed point.   We therefore calculate the fixed point as a function of distance to find the upper bound on the fidelity which can be attained for given $\mathcal{F}_0, L/L_0, p$, and $\eta$.

\subsection{Asymptotic Fidelity}

\begin{figure}
 \vspace{.2in}
\centerline {
\includegraphics[ width=2in]{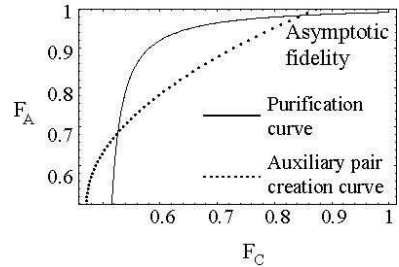}
}
\vspace{.0in}\caption{\it{Approach to asymptotic fidelity.  The solid curve shows the purified fidelity obtained from the auxiliary pair, while the dotted curve corresponds to the auxiliary pair (constructed from two smaller purified pairs) on the next nesting level.  The system moves between the curves at each nesting step, and the upper intercept of the two curves gives the asymptotic fidelity.  For this calculation $F_0 = p = \eta = 0.99, $ and $\upsilon = 0$.}}
%comment: problem with this label (after (54) my version
%gives link to wrong fig. asymp used in label twice, changed the name here.
\label{intersect}
\end{figure}

As the distance increases $L\rightarrow \infty$, the fixed point fidelity can approach an asymptotic value $F_\infty$.   We can understand the existence of $F_{\infty}$ and its value by examining the protocol as a function of nesting level.   
In particular, to generate entanglement over $n$ repeater stations we operate at nesting level $i \sim \log_2{n} $, where we obtain a purified pair 
\beq
\mathcal{F}_A^{(i)} = \mathcal{F}_{FP}(\mathcal{F}_C^{(i)},\eta, p),
\label{fa}
\eeq  
where $\mathcal{F}_{FP}$ is the fixed point solution to Eq.~(\ref{fp}), and $(\mathcal{F}_{FP})_1 = F_{FP}$ is the fixed point fidelity.
We will then use this purified pair $\mathcal{F}_A^{(i)}$ to build up an auxiliary C pair on the next nesting level $i+1$.  Since the fidelity over distance $n-1$ is greater than that over distance $n$, i.e. $(\mathcal{F}_A(n-1))_1 \gtrsim (\mathcal{F}_A(n))_1$, the auxiliary pair fidelity we obtain will be greater than or equal to the first component of 
\beq
\mathcal{F}_C^{(i+1)} \sim C(\{\mathcal{F}_A^{(i)}, \mathcal{F}_A^{(i)}, \mathcal{F}_0, \mathcal{F}_0, \mathcal{F}_0\},\eta, p),
\label{fc}
\eeq
where $C$ is again the connection function.  This auxiliary pair will
then determine  $\mathcal{F}_A^{(i+1)} =
\mathcal{F}_{FP}(\mathcal{F}_C^{(i+1)},\eta, p)$.   When
$\mathcal{F}_A^{(i+1)} = \mathcal{F}_A^{(i)}$, we have reached the
asymptotic fidelity $F_{\infty}=(\mathcal{F}_A)_1$ (see
Figure~\ref{intersect}), % changed name in reference and in a
%label below (asymp was
%used twice) I think this is the way it should but you better check Lily!
which is given by the intersection of the
purification curve Eq.~(\ref{fa}) and the auxiliary pair creation
curve Eq.~(\ref{fc}).   
%The purification curve $\mathcal{F}_A(\mathcal{F}_C)$ and auxiliary
%pair creation curve $\mathcal{F}_C^{-1}(\mathcal{F}_A)$ reveal
%whether or not the fidelity will asymptote at long distances, and
%what that asymptote will be (see Figure~\ref{asymp}).   The system
%moves between the curves at each nesting step, and the intercept of
%the two curves gives the asymptotic fidelity.   When the local errors
%or initial infidelity are too high, the two curves do not intersect,
%and the fidelity decreases monotonically with distance.     

As was the case for the fixed point analysis, we must account for all diagonal components of the density matrix in the Bell state basis (not just the fidelity $a$).   Consequently the asymptotic fidelity represents an upper bound to which the system may converge in the manner indicated by our simulations.
 Finally, we should stress that our calculations have not incorporated loss due to the long but finite memory time in the nuclear spins.   This loss increases with the total time required for repeater operation, and sets the upper limit on the distance over which our scheme could operate.  

\subsection{Results}
%In Figure 1c we present a method which accomplishes this with minimal physical resources at each node.  It requires generating an auxiliary entangled electron spin pair (which we label ``$C$") to purify an entangled pair stored in the nuclear degrees of freedom (``$B$"). %as illustrated in Figure~1c.    The purification protocol described in ~\cite{Deutsch96,Dur99} is then performed by entangling the electron and nuclear spins  via a  hyperfine interaction  at each node% (which is analogous to a two-qubit gate in the
%nuclear-electron spin basis) , and subsequently measuring the electron spins. %, following the protocol described in %  At each purification
%step we employ a protocol 
% that is particularly effective in reducing  phase errors. Comparison of the measurement outcomes  reveals whether the purification step was successful, resulting in a new stored pair $B$ with higher fidelity.  % If the purification step was successful we may repeat the procedure.  After successfully repeating the procedure for $m$ purification steps, the stored pair becomes a purified (``A") pair, which can then be used to create B and C pairs over longer distances.  This procedure is analogous to a pioneering proposal~\cite{briegel98}; however, by incorporating two extra connection steps in generating the auxiliary $C$ pair,  the required physical resources are reduced, so that the protocol can be implemented with a single electronic and nuclear spin at each node.

\begin{figure}\vspace{.0in}
\centerline {
\includegraphics[ width=2.5in]{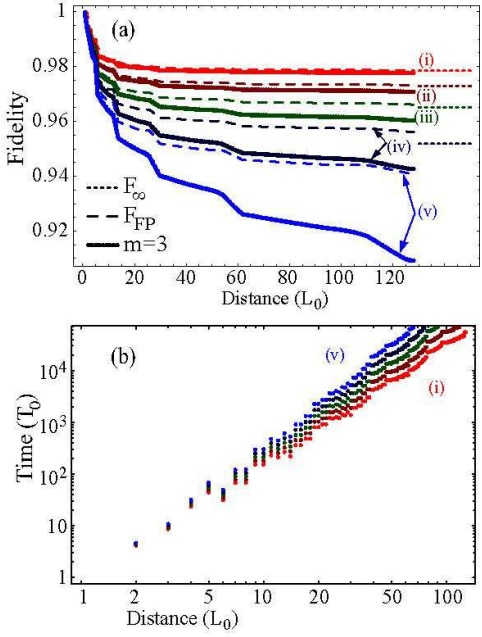}
}
\vspace{.0in}\caption{%Scaling of quantum communication fidelity. 
\it{ (a) Fidelity scaling with distance.   Points show results using 3 purification steps at each nesting level;  dashed lines show the fixed point $F_{FP}$ at each distance; dotted lines indicate the
asymptotic fidelity $F_{\infty}$.   For (a) and (b), measurements and local two-qubit operations $\eta=p$ contain $0.5\%$ errors.  The initial fidelity $F_0$ is (i) 100\% (ii) 99\% (iii) 98\% (iv)
97\% (v) 96\% with phase errors only.   (b) Average time scaling with distance
for m=3, given in units of $T_0 = (t_0 + t_c)/P$, the time required to
generate entanglement between nearest neighbors,  and $L_0$, the
distance between nearest neighbors.   Measurement and local operation
times are neglected.
Note that the axes are logarithmic, so time scales polynomially with
distance. }
}  
\label{asymp}
\end{figure}

The discussion of final fidelity may be summarized as follows: the fidelity obtained at the end of this nested purification procedure, $F(m, L, F_0, p, \eta)$, depends on the number of purification steps $m$, the distance $L$ between the outer nodes, the initial fidelity $F_0$ between adjacent nodes, and the reliability of measurements $\eta\leq1$ and local two-qubit operations $p\leq1$  required for entanglement purification and  connection~\cite{Dur99}.   As the number of purification steps increases  $m\rightarrow \infty$, the fidelity at a given distance $L$ approaches a fixed point $F\rightarrow F _{FP}(L, F_0, p, \eta)$ at which additional purification steps yield no further benefit~\cite{Dur99}.  Finally, as $L$ increases,  the fidelity may approaches an asymptotic value $F_{FP}\rightarrow F_{\infty}(F_0, p, \eta)$.
 Figure~\ref{asymp}a illustrates the efficiency of the purification protocol:
 for initial fidelities $F_0\gtrsim 97\%$, three purification steps
 suffice to produce entanglement at large distances.

Figure~\ref{asymp}b demonstrates that our scheme permits generation of high-fidelity, long distance entangled pairs in the presence of percent-level errors in polynomial time. 
Because solid-state devices allow fast operations and measurements,
the overall time scale is set by the classical communication time
between nodes.  %%%%Lily which collection and detection efficiency did
%%%%you use?
As an example,  using a photon loss rate of $\sim 0.2$ dB/km and inter-node separation $L_0 \sim 20$ km (so that in the limit of good detectors the collection efficiency is $ 10^{-0.4} \sim 1/e$),  a fidelity set by an emission probability $P_{\rm{em}} \sim 8\%$, local errors $\eta = p = 0.5\%$, and just one purification step at each nesting level, our scheme could potentially produce entangled pairs with fidelity $F\sim 0.8$ sufficient to violate Bell's inequalities over 1000 km in a few seconds.    For comparison, under the same set of assumptions direct entanglement schemes would require $\sim10^{10}$ years.   %Moreover, it is likely that the bit-rate could be significantly improved by employing optimal control theory to tailor the details of the repeater protocol to the parameters of a desired implementation. %, for instance introducing a variable number of purification steps or degree of parallelism at each nesting level.  

\begin{figure}\vspace{.0in}
\centerline {
\includegraphics[ width=3.5in]{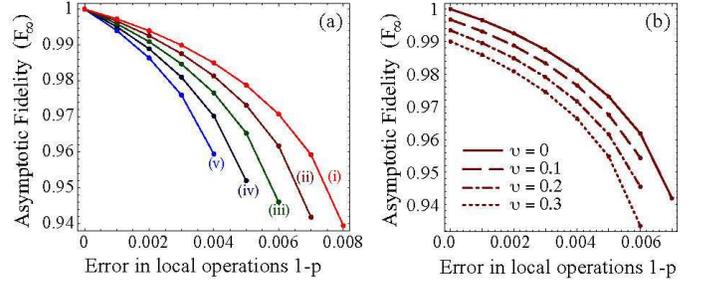}
}
\vspace{.0in}\caption{%Scaling of quantum communication fidelity. 
\it{(a) Long-distance asymptote dependence on initial fidelity  $F_0$ of (i) 100\% (ii) 99\% (iii) 98\% (iv)
97\% (v) 96\% with phase errors only.  (b) Long-distance asymptote dependence on error type.   For the calculations shown,  $F_0 = 0.99$,  and the shape parameter ranges from $\upsilon = 0$ to $\upsilon = 0.3$.  In both (a) and (b) measurement errors are set equal to operational errors, $\eta = p$.  % (solid), $\upsilon = 0.1$  (dashed), $\upsilon = 0.2$ (dash-dotted), and $\upsilon = 0.3$ (dotted).
}
} 
\label{asymp2}\end{figure}

Fig.~\ref{asymp2}a  shows  
that our scheme will operate in the presence of $1-p \lsim 1\%$ errors
in local operations and percent-level phase errors in initial
entanglement fidelity.    
Other types of error are in principle possible, and we consider nonzero shape parameters $\upsilon$ for the initial fidelity $\mathcal{F}_0$ in Eq.~(\ref{shape}).    The asymptotic
fidelity shown in 
Fig.~\ref{asymp2}b  indicates that,
 although the protocol we use is most effective for purifying phase
 errors, it also tolerates arbitrary errors. 

\subsection{Optimization}

Once the parameters of the system are established, the protocol can be
optimised to minimise the time required to generate some minimum
fidelity $F_{min}$ over a distance $L$.   We can vary the number of
repeater stations $\sim L/L_0$ and the number of purification steps
$m$ (which need not be constant).  We can also tailor the entanglement
generation procedure by changing the emission probability $P_{em}$ to
find the optimum balance between initial infidelity $1-F_0\sim P_{em}$
and entanglement generation time $T_0 \propto 1/P_{em}$.   Finally,
one could use more advanced optimal control techniques to vary the
details of the protocol itself.   In particular, it should be possible
to speed up the algorithm by working simultaneously on multiple
nesting levels, beginning entanglement generation and connection on
the next nesting level as soon as the interior nodes are free.
Further speed-up may also be possible in the case when collection
efficiency is very high by using coincidence detection in combination
with e.g. time-bin encoding~\cite{gisin02}. 
As noted previously such coincidence detection could also be
advantageous for interferometric stability \cite{simon,anders98,vanenk97}.

Ultimately, the speed of this protocol is limited by three factors:  classical communication time between nodes, probabilistic entanglement generation, and sequential purification.   Faster techniques will require more efficient entanglement generation or larger numbers of qubits at each node to allow simultaneous purification steps.  

\subsection{Comparison to other quantum repeater schemes}

This scheme combines the advantages of two pioneering proposals for
quantum repeaters \cite{briegel98, Duan01}.    Early work showed that
entanglement purification and swapping could be combined to permit
efficient, fault-tolerant quantum communication over distances longer
than an attenuation length \cite{briegel98}.   This scheme
incorporated error correction at the cost of increased physical
resources, requiring nodes containing a number of qubits scaling at
minimum logarithmically with distance \cite{Dur99}.  Owing to the
difficulty of implementing even few-qubit quantum computation,
implementation of this scheme remains a challenging goal.   Our scheme
is closely related to the 
%Anders changed from Innsbruck to original (we confuse the reader by
%saying Innsbruck when we haven't done it before)
original proposal with one key difference:
by spatially rearranging the required physical resources, we can
efficiently simulate their protocol while maintaining a constant
requirement on qubits per node.   This makes our scheme amenable to
realistic physical implementation. 

Another  physical implementation for quantum
repeaters uses atomic ensembles as a long-lived memory for
photons\cite{Duan01}.   Entanglement is generated by interfering Raman
scattered light from two ensembles.  The entanglement is
probabilistically swapped using an EIT readout technique.  This scheme
elegantly avoids effects of the dominant photon loss error by
conditioning success on photon detection.    Our scheme primarily
differs 
%anders changed from Duan to this, same reson as above
from this proposal in two ways:  first, access  to two-qubit
operations between electron and nuclear spin permits deterministic
entanglement swapping; second, the two-qubit nodes allow active
correction of arbitrary errors.   

\section{Physical systems}

We conclude with three specific examples for potential implementation of
the presented method.   

\subsection{Implementation with NV centers}

%The nitrogen vacancy (NV) center in diamond has a strong, state-selective optical transition (Figure~3a) near 637nm which has been used for robust generation of  single photons on demand ~\cite{weinfurter00,beveratos02} and single spin measurement ~\cite{jelezko04}.  The  triplet electron spin ground state is strongly coupled to a  nearby $^{13}$C impurity nuclear spin, which can have a very long  coherence time~\cite{Ramanathan04}.  Spin selective fluorescence allows electron spin initialization, measurement~\cite{jelezko04} and entanglement with outgoing photons;  electron spin resonance (ESR) and nuclear magnetic resonance (NMR) have already been employed to manipulate coupled electron and nuclear spins~\cite{jelezko04b}.  In the ground state, the energy splitting between electron spin states $M_s = 0$ and $M_s = \pm 1$ is an order of magnitude larger than the hyperfine interaction, effectively decoupling the nuclear and electronic spin states.  Since  the optical transition frequency between states with $M_s=0$ is independent of nuclear spin state,  information may be safely stored in the nuclear spin during light scattering. 
The NV center level structure illustrated in Fig.~\ref{shelf} allows implementation of all steps in the repeater protocol.  The cycling transition from $\ket{0}$ to $\ket{e}$ is used for electron spin initialization by measurement, entanglement generation, and electron spin state measurement.   A series of ESR and NMR pulses can be used to perform arbitrary gates between the electron spin and an adjacent $^{13}$C spin \cite{jelezko04b}.  Consequently, nuclear spin state initialization and  measurement is achieved by initializing the electron spin, mapping the nuclear spin state onto the electron spin, and subsequently measuring the electron spin.  
Entanglement propagation and purification can be implemented in NV
centers by driving ESR and NMR transitions and using optical detection
of the electron spin states.  Once electron spin entanglement is
established between nodes $R_i$ and $R_{i-1}$, it can be transferred
to the nuclear spins, leaving the electron degree of freedom 
free
to generate entanglement between station $R_i$ and $R_{i+1}$.
Provided that we can reinitialize the electron spin without 
 affecting the
nuclear entanglement, we can perform the same probabilistic
entanglement procedure.   Note that ESR multiplexing is required to
perform a $\pi/2$ pulse independent  of the nuclear spin;  this can be
accomplished simply by applying two ESR pulses at the two transition
frequencies.   

We now consider the feasibility of implementing our repeater protocol using NV centers in diamond.  %lily added:
Owing to the overlap of electron wavefunctions in the ground and excited states, most of the NV center optical emission goes into the phonon sidebands.   
Other color centers in diamond, for example the NE8 center \cite{Gaebel04, Kilin}  may  suffer less from this drawback.    To enhance the relative strength of the zero-phonon line, it will be necessary to couple the NV center to a cavity.  
%end of lily addition 
For NV centers coupled to cavities with Purcell factors  $\sim
10$~\cite{santori02}, we find that the dominant source of error is electron spin
decoherence during the classical communication period.  Using an
emission probability $P_{\rm{em}}\sim 5\%$, a collection efficiency
$\epsilon \sim 0.2$, and a  classical communication time of $t_c \sim
70\mu$s over $L_0 \sim 20$ km, we find the fidelity of directly
entangled pairs can reach $F_0 \sim 97\%$ for electron spin coherence
times in the range of a few milliseconds.   Electron spin coherence
times in the range of $100\mu$s have been observed at room temperature
 and significant improvements are expected for
 high purity samples at low temperatures~\cite{Kennedy03}.  The large hyperfine splitting allows fast local operations between electron and nuclear spin degrees of freedom on a timescale $\sim 100$ns~\cite{jelezko04b} much
shorter than the decoherence time, allowing $1-p < 1\%$. Finally,  cavity
enhanced collection should significantly improve observed measurement
efficiencies of $\eta \sim 80\%$~\cite{jelezko04b}.

\subsection{Alternative implementation: quantum dots}

Our discussion thus far has attempted to remain general while exemplifying our proposal using NV centers.   The basic idea of using two-qubit repeater stations should be applicable to a wide variety of systems featuring coupled electron and nuclear spins.   To illustrate an alternative implementation, we consider doped self-assembled quantum dots whose electron spin is coupled to collective nuclear states in the lattice.  Compared to NV centers, this system offers large oscillator strengths and the potential for Raman manipulation.  
Doped semiconductor quantum dots have been considered in a variety of quantum computing proposals and related technologies~\cite{imamoglu99,pazy03}.  The spin state of the dopant electron provides a natural qubit with relatively long coherence times.  Assuming a high degree of nuclear spin
polarization ($P_{\rm n} \gtrsim 0.95$)~\cite{bracker04} and active ESR
pulse correction,  the electron spin dephasing time is expected to be $1$
ms~\cite{golovach03}.  The spins of lattice nuclei in the quantum dot
provide an additional, quasi-bosonic degree of freedom with extremely
long coherence times ($\sim 1$ s with active
correction~\cite{Ramanathan04}).  Such ensembles of nuclear spin have
been considered for use as a quantum memory~\cite{taylor03} and, by
taking advantage of the non-linearity of the Jaynes-Cummings
Hamiltonian, 
as a fundamental qubit for a quantum computer~\cite{taylor04}.

\begin{figure}\vspace{.0in}
\centerline {
\includegraphics[ width=3.5in]{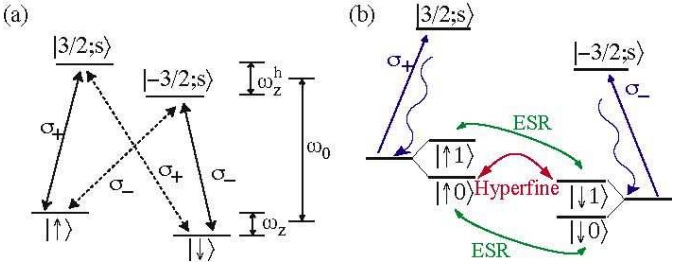}
}
\vspace{.0in}\caption{\it{(a) Level structure for single electron to trion transition in a single-electron doped III-V or II-VI quantum dot with external magnetic field in close to a Faraday geometry with Zeeman splitting $\omega_z$, heavy-hole splitting $\omega^h_z$ and up to four optical fields of different frequency and polarization.  Dashed lines indicate weak dipole moments due to small magnetic field mixing.  The triplet two-electron states are not included due to a large ($> 1 $meV) exchange energy allowing for complete suppression of their effects.
(b) Electronic ($\ket{\uparrow}, \ket{\downarrow}$) and collective
    nuclear ($\ket{0}, \ket{1}$) states and their transitions for
    singly doped quantum dots in a polarized nuclear spin lattice.
 } }
\label{qd}\end{figure}

Unlike the spin triplet state of the NV centers, the conduction band electron has two states, $\ket{\uparrow}$ and $\ket{\downarrow}$,  corresponding to spin aligned and anti-aligned with an external magnetic field $B_{ext} || \hat{z}$.   The quantum dot system also differs from NV centers in that it can be manipulated using Raman transitions: when the external field and growth direction are perpendicular (Voigt geometry), two allowed optical transitions to a trion state produce %lily removed "s" 
 a lambda system %via two optical fields at frequencies $\nu_{\pm} = \nu \pm \omega_{\rm Zeeman}/2$ 
; moving towards aligned field and growth directions (Faraday geometry) suppresses the``forbidden'' transitions, as shown in Fig.~\ref{qd}a.  %%DO WE HAVE A REFERENCE FOR RAMAN ESR??%%

%\begin{figure}
%\includegraphics[width = 2 in]{levelstruct080804.eps}
%\caption{\it{
%Level structure for single electron to trion transition in a single-electron doped III-V or II-VI quantum dot with external magnetic field in close to a Faraday geometry with Zeeman splitting $\omega_z$, heavy-hole splitting $\omega^h_z$ and up to four optical fields of different frequency and polarization.  Dashed lines indicate weak dipole moments due to small magnetic field mixing.  The triplet two-electron states are not included due to a large ($> 1 $meV) exchange energy allowing for complete suppression of their effects.
%}}
%\label{qdold}
%\end{figure}

Electron spin coherence can thus be prepared via Raman transitions or by standard ESR setups, and changes in effective magnetic field can be accomplished by off-resonant, spin-dependent AC Stark shifts with $\sigma_+$ light.

Although optical transitions in doped quantum dots can exhibit  homogeneous broadening $\Gamma  \sim 100$ GHz$\sim 10-100\gamma$~\cite{kiraz02},  the corresponding error  can be made negligible by sending the output from the cavity through a frequency filter with a linewidth of a few hundred MHz.  \endnote{For our entanglement generation scheme,  such a filter will allow the desired narrow band of coherent light to pass through while rejecting the broad incoherent background.  Consequently the filter will not decrease collection efficiency in the desired mode.}  Moreover, we note that InAs quantum dots have been successfully coupled to  microcavities with Purcell factors $\sim 10$~\cite{santori02}.  

%lily combined two quantum dot figures

Whereas the NV center electron spin was coupled to a single nuclear impurity, the electron in a quantum dot couples to collective excitations of many thousands of nuclei.   %lily added here to orient the reader
We briefly discuss this system; further details are given in Refs.~\cite{taylor03, taylor03b}.
The Hamiltonian governing this interaction is
\beq
H_{qd} = \omega_z \Sz + \hbar  \sum_k \gamma_k \Iz^k + \hbar \Omega \sum_k \lambda_k \SV \cdot \IV^k,
\eeq
where $\gamma_k$ is the gyromagnetic ratio for nuclear spin $\IV^k$, the nuclear spin coupling amplitudes satisfy $\lambda_k \propto |\psi(r_k)|^2$, $\sum_k \lambda_k^2 = 1$, and $\Omega = A / \hbar \sum_k \lambda_k$ ($A$ is the hyperfine interaction constant).    By identifying collective nuclear spin operators, $\AV = \sum_k \lambda_k \IV$, the hyperfine term may be written $\hbar \Omega \SV \cdot \AV$.  For simplicity we restrict the following discussion to the case of perfect nuclear polarization (so the initial state of all $N_n$ nuclear spins in the quantum dot is $\ket{0} = \ket{-I}\otimes \ldots \otimes \ket{-I}$ for $I$-spins).  Then an effective Jaynes-Cummings type Hamiltonian describes the system:
\beq
H_{qd}^{eff} = \hbar \omega_z^{eff} \Sz + \hbar \Omega/2 (\Ap \Sm + \Am \Sp) .
\eeq
with corrections of order $\Omega / \sum_k \lambda_k \sim A/N_n$.
The effective Zeeman splitting  $\omega_z^{eff}=\omega_z - I A/\hbar$ %, produced by the external magnetic fields and $\Sz \Az$ terms, 
 is  dominated by the field associated with the polarized nuclear spins %, where the nuclear state shifts the effective field by $I A/\hbar$
; for example, in GaAs quantum dots, this Overhauser shift is $I A/\hbar \sim 33$GHz.   The large detuning $\omega_z^{eff}$ suppresses interactions which exchange energy between the electron and nuclear spins.   By changing the effective magnetic field, we can shift the system into resonance $\omega_z^{eff}\rightarrow 0$ to drive Rabi oscillations between the electron spin and the collective nuclear state,   
%Defining collective nuclear spin excitations as $\ket{m} \propto (\Ap)^m \ket{0}$, each subspace of excitation number $m$ is spanned by $\{ \ket{m} \ket{\downn},\ket{m-1}\ket{\upp} \}$.  Dynamics between hyperfine levels are driven by the off-axis (xy-terms) when the effective Zeeman splitting is small $|\omega_z^{eff}| \ll \Omega$.  
see figure \ref{qd}b. 
Pulsing the appropriate effective field permits a controllable map between the electron spin state and the collective nuclear degrees of freedom spanned by $\ket{0}$ and $\ket{1} = \Ap \ket{0}$.   %by starting in the state $(\alpha \ket{\upp} + \beta \ket{\downn}) \ket{0}$ and at $\omega_z^{eff} = 0$, waiting a time $\pi/\Omega$, at which point the system evolves to
%$\ket{\downn} (i \alpha \ket{1} + \beta \ket{0})$.  This is described by the unitary $\hat U_{map}$. 
 In addition, more complicated sequences of electron-nuclear spin interaction and electron spin manipulation allow for arbitrary two qubit operations on $\ket{\upp},\ket{\downn}$ and $\ket{0},\ket{1}$ (see Ref.~\cite{taylor04} and commentary therein).

Measurement and initialization proceed in much the same manner as described for NV centers.  The state of the electron spin system can be read out by exciting a cycling transition with resonant $\sigma_+$ light, and measurement and ESR (or Raman transitions) can be employed to initialize the system in the desired state.  As the effective Knight shift of the electron spin is negligible on the time scales of entanglement preparation, the collective nuclear state's coherence is unaffected by this process.  Due to the improved selection rules and possibility of Raman transitions, it may be more effective to use the Raman entanglement generation scheme.

The nuclear state of the quantum dot can be prepared by cooling the nuclear spins using preparation of electron spin and manipulation of the effective magnetic field~\cite{imamoglu03}.  In practice, this leaves the nuclear system in a state $\ket{\mathcal D}$ with the same symmetry properties as the state $\ket{0}$ described above~\cite{taylor03b}.   To date, 60\% nuclear spin polarization has been achieved by optical pumping in GaAs quantum dots~\cite{bracker04}.   As was the case with NV centers, the nuclear spin state can be read out by preparing the electron spin in the $\ket{\downarrow}$ state, mapping the nuclear state to the electron spin state%via $\hat U_{map}^{\dag}$
, and measuring the electron spin state.  
 
 \subsection{Atomic Physics Implementation}

Compared to the solid state implementations we have considered so far,
implementations in single trapped atoms or ions have the advantage that they
typically have 
very 
little broadening of the optical transitions. %In essence,  lily  changed this wording
Because atomic systems do %es 
not reside in a complicated many-body
environment, % and 
their internal degrees of freedom %of the atoms can therefore 
can have very long coherence times. For most atomic systems, however,
it is %, lily moved however,  
hard to identify a mechanism which allows %for 
one degree of freedom, e.g., the
nuclear spin, to be decoupled while we probe some other degree of
freedom, e.g., the electron spin. 
Below we describe a system which
does fulfill this
requirement, although practical considerations indicate implementation may be challenging.
% JMT:  removed (although it will be  challenging to implement in practice).

\begin{figure}[htbp]

\centerline {
\includegraphics[width=2in]{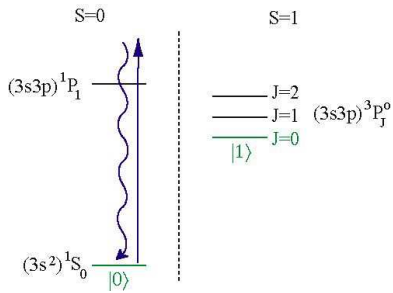}
}  
  \caption{\it{Electronic level structure of atomic magnesium. The
  electronic ground 
  state $^1S_0$ has
  vanishing spin and orbital angular momentum.  In this state the
  nuclear spin therefore decouples from the electronic degrees of
  freedom.   An entangling operation which is also insensitive to the
  nuclear degree of freedom can be achieved with lasers which are
  detuned much further than the hyperfine splitting in the $^1P_1$
  level.}}
  \label{fig:mg}
\end{figure}

We consider alkali-earth atoms, such as neutral magnesium, and chose an isotope
with non-vanishing nuclear spin ($^{25}$Mg). The lowest lying  states of
magnesium %is 
are shown in Fig.~\ref{fig:mg} (electronic structure only). Instead
of the electronic spin states we have considered so far, i.e., for NV centers and quantum dots,  we will use
states which differ both in spin and orbital angular momentum. The stable ground state $^1S_0$ will
serve as state $|0\rangle$.   In this state, the electronic degrees of freedom have neither spin nor orbital angular momentum  and the nuclear spin is thus decoupled from the electronic
state.  The excited state $^3P_0^o$ (whose hyperfine interactions also vanish to leading order) will provide state $|1\rangle$.  Note that the triplet-singlet transition from  $^3P_0^o$ to the
ground state is highly forbidden and this state has an
extremely long lifetime,   %Anders added
but transitions between the two states can still be induced with a strong
laser.

To create entanglement we couple the ground state to the excited state
$^1P_1$ with a laser field and collect the scattered light. From
this excited state the atom essentially always decays back into the
ground state. If the
driving is detuned much further than the hyperfine splitting in the
excited %lily added 
state,  the nuclear spin is also decoupled during this process. 
The nuclear spin can therefore by used to store information 
while we entangle the electronic state with another atom. Finally, to implement
gates between the electronic and nuclear states one should, for
instance, couple   %for instance 
the $|0\rangle$ state to another state in the atom where there is a hyperfine
interaction, %lily added
for example using resonant excitation of the $^1P_1$ state.
%JMT: which state?  Can we be more specific?

%Lily added:
Finally, we note that all three physical implementations we suggest operate in the visible or near-IR, and will likely require high-efficiency frequency conversion to telecom wavelengths for low-loss photon transmission.  

\section{Conclusion}

In conclusion, we propose a method for fault tolerant quantum
communication over long distances requiring only probabilistic
nearest-neighbor entanglement generation%jmt added
, two-qubits per node, 
and two-qubit operations.  We
compare several schemes for entanglement generation and discuss two
solid-state systems %anders added
and an atomic system which might be used to implement them.  Potential
applications %jmt cleaned: may
include
%, for instance, 
secure transmission of secret
messages over intercontinental distances.

The authors wish to thank Phillip Hemmer, Aryesh Mukherjee,  %anders added
Klaus M\o lmer,
      Alexander Zibrov, and Gurudev Dutt. %for useful conversations.
      This work is supported by DARPA, NSF, ARO-MURI, and the Packard, Sloan and Hertz Foundations and  the Danish Natural Science Research Council.

\section{Appendix}
Photo-bleaching is a detectable error, so it does not affect the
fidelity of entanglement generation or measurement, as we described above.
However, it can
increase the time required for these operations.    Fluorescence
correlation experiments are consistent with assigning a metastable
singlet structure to the shelving state, which is coupled strongly to
the $M_s = \pm 1$ excited states but only weakly to the $M_s = 0$
excited state (see Figure~\ref{shelf}), \cite{nizovtzev03}.   We need to account for the
possibility that our NV center bleaches during entanglement
generation, requiring us to start over.   During each attempt we
resonantly excite the $M_s = 0$ transition with some probability 
% Anders inserted \rm
$P_{{\rm em}}$ (see Eq.~\ref{Pem}).    To quantify the population lost to
$\ket{W}$ we will consider a model where the $M_s = \pm1$ excited
states decay to the shelving state at rate $\Gamma_S$.  The oscillator
strength for the $M_s = \pm1$ optical transitions are unknown, so we
will assume that the Rabi frequencies on the $M_s = \pm 1$ transitions
are $\Omega' $.   

During one attempt at entanglement generation, the probability to end up in the shelving state is 
\beq
P_W \sim \Gamma_S\frac{\Omega'^2}{\delta^2} t_0,
\eeq
where $\delta$ is the detuning from the $M_s = \pm1$ optical
transition.  (The excited state energies are strongly inhomogeneously
broadened, so $\delta$ is not precisely known; this detuning should be
controllable using strain or applied electric fields.)  On average, a
large number of attempts   %anders insrted rm and changed from ex
$\sim 4/P_{{\rm em}}\epsilon^2$
%\beq
%n_0 \approx \frac{4}{P_{em}\epsilon^2} \approx \frac{4 (\gamma + \Gamma)}{\Omega^2 t_0 \epsilon^2}
%\eeq
are required for successful entanglement generation.  Consequently the total probability for the system to end up in the shelving state during entanglement generation is
\beq
P_W \sim 4 \Gamma_S\frac{\Omega'^2}{\delta^2} \frac{\gamma}{\Omega^2  \epsilon^2} \sim 4 \frac{\Gamma_S \gamma}{\delta^2 \epsilon^2}\frac{\mu'^2}{\mu^2},
\eeq
where $\mu (\mu')$ is the oscillator strength for the $M_s = 0(\pm1)$ transition.
The precise values of these parameters are unknown, but we can estimate their order of magnitude: $\Gamma_S \sim 1-10$ MHz, $\gamma +\Gamma \sim 100$ MHz, $\delta \sim 1$ GHz, $\mu'\sim\mu$, yielding
\beq
P_W \sim \frac{4 (10^{-3}-10^{-4})}{\epsilon^2}.
\eeq
If this error rate is too large, we can also check for photo-bleaching at intervals during the entanglement procedure.
%This calculation indicates that high collection efficiency $\epsilon^2$ will be critical for reducing deleterious effects of the shelving state.

The shelving state poses a similar problem during measurement.  In
this case, the $M_s = 0$ transition is strongly illuminated so that at
least one photon reaches the detectors:
%anders changed ex to em
$P_{{\rm em}}\sim 1/\epsilon^2$.      Under the same illumination, any
population in $\ket{1}$ will end up in the shelving state with
probability %anders changed ex to em
\beq
P_W \sim P_{{\rm ex}} \frac{\Gamma
\gamma}{\delta^2}\frac{\mu'^2}{\mu^2} \sim \frac{\Gamma
\gamma}{\delta^2 \epsilon^2}\frac{\mu'^2}{\mu^2} \sim
\frac{(10^{-3}-10^{-4})}{\epsilon^2}, 
\eeq
%which again emphasizes the importance of high collection efficiency in obtaining a successful measurement.   
Note that the measurement fidelity is unaffected by photo-bleaching if we verify that the center is optically active by observing fluorescence either directly  from the $\ket{0}$ state or after applying a multiplexed ESR pulse to the $\ket{1}$ states.
Ultimately, in this model the only effect of the shelving state is to reduce the success rate for entanglement generation and measurement by of order a few percent.   Finally, we should note that the effect of the shelving state on the nuclear spin state is currently not known, and could potentially complicate the sequence of operations necessary upon detection of a shelving event.  

\bibliography{repeat_prl2}

\begin{thebibliography}{48}
\expandafter\ifx\csname natexlab\endcsname\relax\def\natexlab#1{#1}\fi
\expandafter\ifx\csname bibnamefont\endcsname\relax
  \def\bibnamefont#1{#1}\fi
\expandafter\ifx\csname bibfnamefont\endcsname\relax
  \def\bibfnamefont#1{#1}\fi
\expandafter\ifx\csname citenamefont\endcsname\relax
  \def\citenamefont#1{#1}\fi
\expandafter\ifx\csname url\endcsname\relax
  \def\url#1{\texttt{#1}}\fi
\expandafter\ifx\csname urlprefix\endcsname\relax\def\urlprefix{URL }\fi
\providecommand{\bibinfo}[2]{#2}
\providecommand{\eprint}[2][]{\url{#2}}

\bibitem[{\citenamefont{Gisin et~al.}(2002)\citenamefont{Gisin, Riborty,
  Tittel, and Zbinden}}]{gisin02}
\bibinfo{author}{\bibfnamefont{N.}~\bibnamefont{Gisin}},
  \bibinfo{author}{\bibfnamefont{G.}~\bibnamefont{Riborty}},
  \bibinfo{author}{\bibfnamefont{W.}~\bibnamefont{Tittel}}, \bibnamefont{and}
  \bibinfo{author}{\bibfnamefont{H.}~\bibnamefont{Zbinden}},
  \bibinfo{journal}{Rev. Mod. Phys} \textbf{\bibinfo{volume}{74}},
  \bibinfo{pages}{145} (\bibinfo{year}{2002}).

\bibitem[{\citenamefont{Brassard et~al.}(2000)\citenamefont{Brassard,
  Lutkenhaus, More, and Sanders}}]{brassard00}
\bibinfo{author}{\bibfnamefont{G.}~\bibnamefont{Brassard}},
  \bibinfo{author}{\bibfnamefont{N.}~\bibnamefont{Lutkenhaus}},
  \bibinfo{author}{\bibfnamefont{T.}~\bibnamefont{More}}, \bibnamefont{and}
  \bibinfo{author}{\bibfnamefont{B.}~\bibnamefont{Sanders}},
  \bibinfo{journal}{Phys. Rev. Lett.} \textbf{\bibinfo{volume}{85}},
  \bibinfo{pages}{1330} (\bibinfo{year}{2000}).

\bibitem[{\citenamefont{Briegel et~al.}(1998)\citenamefont{Briegel, Dur, Cirac,
  and Zoller}}]{briegel98}
\bibinfo{author}{\bibfnamefont{H.~J.} \bibnamefont{Briegel}},
  \bibinfo{author}{\bibfnamefont{W.}~\bibnamefont{Dur}},
  \bibinfo{author}{\bibfnamefont{J.~I.} \bibnamefont{Cirac}}, \bibnamefont{and}
  \bibinfo{author}{\bibfnamefont{P.}~\bibnamefont{Zoller}},
  \bibinfo{journal}{Phys. Rev. Lett.} \textbf{\bibinfo{volume}{81}},
  \bibinfo{pages}{5932} (\bibinfo{year}{1998}).

\bibitem[{\citenamefont{{Zukowski \it{et al.}}}(1993)}]{zukowski93}
\bibinfo{author}{\bibfnamefont{M.}~\bibnamefont{{Zukowski \it{et al.}}}},
  \bibinfo{journal}{Phys. Rev. Lett.} \textbf{\bibinfo{volume}{71}},
  \bibinfo{pages}{4287} (\bibinfo{year}{1993}).

\bibitem[{\citenamefont{{Bennett \it{et al}}}(1993)}]{bennett93}
\bibinfo{author}{\bibfnamefont{C.}~\bibnamefont{{Bennett \it{et al}}}},
  \bibinfo{journal}{Phys. Rev. Lett.} \textbf{\bibinfo{volume}{70}},
  \bibinfo{pages}{1895} (\bibinfo{year}{1993}).

\bibitem[{\citenamefont{{Bouwmeester \it{et al.}}}(1997)}]{bouwmeester97}
\bibinfo{author}{\bibfnamefont{D.}~\bibnamefont{{Bouwmeester \it{et al.}}}},
  \bibinfo{journal}{Nature} \textbf{\bibinfo{volume}{390}},
  \bibinfo{pages}{575} (\bibinfo{year}{1997}).

\bibitem[{\citenamefont{Ekert}(1991)}]{ekert91}
\bibinfo{author}{\bibfnamefont{A.}~\bibnamefont{Ekert}},
  \bibinfo{journal}{Phys. Rev. Lett.} \textbf{\bibinfo{volume}{67}},
  \bibinfo{pages}{661} (\bibinfo{year}{1991}).

\bibitem[{\citenamefont{{Bennett \it{et al.}}}(1996)}]{Bennett96}
\bibinfo{author}{\bibfnamefont{C.}~\bibnamefont{{Bennett \it{et al.}}}},
  \bibinfo{journal}{Phys. Rev. Lett.} \textbf{\bibinfo{volume}{76}},
  \bibinfo{pages}{722} (\bibinfo{year}{1996}).

\bibitem[{\citenamefont{{Deutsch \it{et al.}}}(1996)}]{Deutsch96}
\bibinfo{author}{\bibfnamefont{D.}~\bibnamefont{{Deutsch \it{et al.}}}},
  \bibinfo{journal}{Phys. Rev. Lett.} \textbf{\bibinfo{volume}{77}},
  \bibinfo{pages}{2818} (\bibinfo{year}{1996}).

\bibitem[{\citenamefont{van Enk et~al.}(1998)\citenamefont{van Enk, Cirac, and
  Zoller}}]{vanenk98}
\bibinfo{author}{\bibfnamefont{S.~J.} \bibnamefont{van Enk}},
  \bibinfo{author}{\bibfnamefont{J.~I.} \bibnamefont{Cirac}}, \bibnamefont{and}
  \bibinfo{author}{\bibfnamefont{P.}~\bibnamefont{Zoller}},
  \bibinfo{journal}{Science} \textbf{\bibinfo{volume}{279}},
  \bibinfo{pages}{205} (\bibinfo{year}{1998}).

\bibitem[{\citenamefont{Duan and Kimble}(2004)}]{duan04}
\bibinfo{author}{\bibfnamefont{L.}~\bibnamefont{Duan}} \bibnamefont{and}
  \bibinfo{author}{\bibfnamefont{H.}~\bibnamefont{Kimble}},
  \bibinfo{journal}{Phys. Rev. Lett.} \textbf{\bibinfo{volume}{92}},
  \bibinfo{pages}{127902} (\bibinfo{year}{2004}).

\bibitem[{\citenamefont{{Blinov \it{et al.}}}(2004)}]{blinov04}
\bibinfo{author}{\bibfnamefont{B.}~\bibnamefont{{Blinov \it{et al.}}}},
  \bibinfo{journal}{Nature} \textbf{\bibinfo{volume}{428}},
  \bibinfo{pages}{153} (\bibinfo{year}{2004}).

\bibitem[{\citenamefont{{Bose \it{et al.}}}(1999)}]{Bose99}
\bibinfo{author}{\bibfnamefont{S.}~\bibnamefont{{Bose \it{et al.}}}},
  \bibinfo{journal}{Phys. Rev. Lett.} \textbf{\bibinfo{volume}{83}},
  \bibinfo{pages}{5158} (\bibinfo{year}{1999}).

\bibitem[{\citenamefont{Ye et~al.}(1999)\citenamefont{Ye, Vernooy, and
  Kimble}}]{Ye99}
\bibinfo{author}{\bibfnamefont{J.}~\bibnamefont{Ye}},
  \bibinfo{author}{\bibfnamefont{D.~W.} \bibnamefont{Vernooy}},
  \bibnamefont{and} \bibinfo{author}{\bibfnamefont{H.~J.}
  \bibnamefont{Kimble}}, \bibinfo{journal}{Phys. Rev. Lett.}
  \textbf{\bibinfo{volume}{83}}, \bibinfo{pages}{4987} (\bibinfo{year}{1999}).

\bibitem[{\citenamefont{Duan et~al.}(2001)\citenamefont{Duan, Lukin, Cirac, and
  Zoller}}]{Duan01}
\bibinfo{author}{\bibfnamefont{L.~M.} \bibnamefont{Duan}},
  \bibinfo{author}{\bibfnamefont{M.~D.} \bibnamefont{Lukin}},
  \bibinfo{author}{\bibfnamefont{J.~I.} \bibnamefont{Cirac}}, \bibnamefont{and}
  \bibinfo{author}{\bibfnamefont{P.}~\bibnamefont{Zoller}},
  \bibinfo{journal}{Nature} \textbf{\bibinfo{volume}{414}},
  \bibinfo{pages}{413} (\bibinfo{year}{2001}).

\bibitem[{\citenamefont{Childress et~al.}(2004)\citenamefont{Childress, Taylor,
  S{\o}rensen, and Lukin}}]{shortpaper}
\bibinfo{author}{\bibfnamefont{L.~I.} \bibnamefont{Childress}},
  \bibinfo{author}{\bibfnamefont{J.~M.} \bibnamefont{Taylor}},
  \bibinfo{author}{\bibfnamefont{A.}~\bibnamefont{S{\o}rensen}},
  \bibnamefont{and} \bibinfo{author}{\bibfnamefont{M.}~\bibnamefont{Lukin}},
  \bibinfo{journal}{submitted to PRL, quant-ph/0410123}
  (\bibinfo{year}{2004}).

\bibitem[{\citenamefont{{McKeever \it{et al.}}}(2004)}]{mckeever04}
\bibinfo{author}{\bibfnamefont{J.}~\bibnamefont{{McKeever \it{et al.}}}},
  \bibinfo{journal}{Science} \textbf{\bibinfo{volume}{303}},
  \bibinfo{pages}{1992} (\bibinfo{year}{2004}).

\bibitem[{\citenamefont{Kurtsiefer et~al.}(2000)\citenamefont{Kurtsiefer,
  Mayer, Zarda, and Weinfurter}}]{weinfurter00}
\bibinfo{author}{\bibfnamefont{C.}~\bibnamefont{Kurtsiefer}},
  \bibinfo{author}{\bibfnamefont{S.}~\bibnamefont{Mayer}},
  \bibinfo{author}{\bibfnamefont{P.}~\bibnamefont{Zarda}}, \bibnamefont{and}
  \bibinfo{author}{\bibfnamefont{H.}~\bibnamefont{Weinfurter}},
  \bibinfo{journal}{Phys. Rev. Lett.} \textbf{\bibinfo{volume}{85}},
  \bibinfo{pages}{290} (\bibinfo{year}{2000}).

\bibitem[{\citenamefont{{Beveratos \it{et al.}}}(2002)}]{beveratos02}
\bibinfo{author}{\bibfnamefont{A.}~\bibnamefont{{Beveratos \it{et al.}}}},
  \bibinfo{journal}{Phys. Rev. Lett.} \textbf{\bibinfo{volume}{89}},
  \bibinfo{pages}{187901} (\bibinfo{year}{2002}).

\bibitem[{\citenamefont{{Michler \it{et al.}}}(2000)}]{michler00}
\bibinfo{author}{\bibfnamefont{P.}~\bibnamefont{{Michler \it{et al.}}}},
  \bibinfo{journal}{Science} \textbf{\bibinfo{volume}{290}},
  \bibinfo{pages}{2282} (\bibinfo{year}{2000}).

\bibitem[{\citenamefont{{Santori \it{et al.}}}(2002)}]{santori02}
\bibinfo{author}{\bibfnamefont{C.}~\bibnamefont{{Santori \it{et al.}}}},
  \bibinfo{journal}{Nature} \textbf{\bibinfo{volume}{419}},
  \bibinfo{pages}{594} (\bibinfo{year}{2002}).

\bibitem[{\citenamefont{Cabrillo et~al.}(1999)\citenamefont{Cabrillo, Cirac,
  Garcia-Fernandez, and Zoller}}]{cabrillo99}
\bibinfo{author}{\bibfnamefont{C.}~\bibnamefont{Cabrillo}},
  \bibinfo{author}{\bibfnamefont{J.}~\bibnamefont{Cirac}},
  \bibinfo{author}{\bibfnamefont{P.}~\bibnamefont{Garcia-Fernandez}},
  \bibnamefont{and} \bibinfo{author}{\bibfnamefont{P.}~\bibnamefont{Zoller}},
  \bibinfo{journal}{Phys. Rev. A} \textbf{\bibinfo{volume}{59}},
  \bibinfo{pages}{1025} (\bibinfo{year}{1999}).

\bibitem[{\citenamefont{Browne et~al.}(2003)\citenamefont{Browne, Plenio, and
  Huelga}}]{browne03}
\bibinfo{author}{\bibfnamefont{D.}~\bibnamefont{Browne}},
  \bibinfo{author}{\bibfnamefont{M.}~\bibnamefont{Plenio}}, \bibnamefont{and}
  \bibinfo{author}{\bibfnamefont{S.}~\bibnamefont{Huelga}},
  \bibinfo{journal}{Phys. Rev. Lett.} \textbf{\bibinfo{volume}{91}},
  \bibinfo{pages}{067901} (\bibinfo{year}{2003}).

\bibitem[{\citenamefont{{Nizovtzev \it{et al.}}}(2003)}]{nizovtzev03}
\bibinfo{author}{\bibfnamefont{A.~P.} \bibnamefont{{Nizovtzev \it{et al.}}}},
  \bibinfo{journal}{Optics and Spectroscopy} \textbf{\bibinfo{volume}{94}},
  \bibinfo{pages}{848} (\bibinfo{year}{2003}).

\bibitem[{\citenamefont{{Muller \it{ et al.}}}(1997)}]{muller97}
\bibinfo{author}{\bibfnamefont{A.}~\bibnamefont{{Muller \it{ et al.}}}},
  \bibinfo{journal}{Appl. Phys. Lett.} \textbf{\bibinfo{volume}{70}},
  \bibinfo{pages}{793} (\bibinfo{year}{1997}).

\bibitem[{\citenamefont{Simon and Irvine}(2003)}]{simon}
\bibinfo{author}{\bibfnamefont{C.}~\bibnamefont{Simon}} \bibnamefont{and}
  \bibinfo{author}{\bibfnamefont{W.}~\bibnamefont{Irvine}},
  \bibinfo{journal}{Phys. Rev. Lett.} \textbf{\bibinfo{volume}{91}},
  \bibinfo{pages}{110405} (\bibinfo{year}{2003}).

\bibitem[{\citenamefont{S{\o}rensen and M{\o}lmer}(1998)}]{anders98}
\bibinfo{author}{\bibfnamefont{A.}~\bibnamefont{S{\o}rensen}} \bibnamefont{and}
  \bibinfo{author}{\bibfnamefont{K.}~\bibnamefont{M{\o}lmer}},
  \bibinfo{journal}{Phys. Rev. A} \textbf{\bibinfo{volume}{58}},
  \bibinfo{pages}{2745} (\bibinfo{year}{1998}).

\bibitem[{\citenamefont{van Enk et~al.}(1997)\citenamefont{van Enk, Cirac, and
  Zoller}}]{vanenk97}
\bibinfo{author}{\bibfnamefont{S.}~\bibnamefont{van Enk}},
  \bibinfo{author}{\bibfnamefont{J.}~\bibnamefont{Cirac}}, \bibnamefont{and}
  \bibinfo{author}{\bibfnamefont{P.}~\bibnamefont{Zoller}},
  \bibinfo{journal}{Phys. Rev. Lett.} \textbf{\bibinfo{volume}{78}},
  \bibinfo{pages}{4293} (\bibinfo{year}{1997}).

\bibitem[{\citenamefont{M{\o}lmer et~al.}(1993)\citenamefont{M{\o}lmer, Castin,
  and Dalibard}}]{jump}
\bibinfo{author}{\bibfnamefont{K.}~\bibnamefont{M{\o}lmer}},
  \bibinfo{author}{\bibfnamefont{Y.}~\bibnamefont{Castin}}, \bibnamefont{and}
  \bibinfo{author}{\bibfnamefont{J.}~\bibnamefont{Dalibard}},
  \bibinfo{journal}{J. Opt. Soc. Am. B} \textbf{\bibinfo{volume}{10}},
  \bibinfo{pages}{524} (\bibinfo{year}{1993}).

\bibitem[{\citenamefont{S{\o}rensen and M{\o}lmer}(2003)}]{anders02}
\bibinfo{author}{\bibfnamefont{A.}~\bibnamefont{S{\o}rensen}} \bibnamefont{and}
  \bibinfo{author}{\bibfnamefont{K.}~\bibnamefont{M{\o}lmer}},
  \bibinfo{journal}{Phys. Rev. Lett.} \textbf{\bibinfo{volume}{90}},
  \bibinfo{pages}{127903} (\bibinfo{year}{2003}).

\bibitem[{\citenamefont{Barrett and Kok}(2004)}]{barrett04}
\bibinfo{author}{\bibfnamefont{S.}~\bibnamefont{Barrett}} \bibnamefont{and}
  \bibinfo{author}{\bibfnamefont{P.}~\bibnamefont{Kok}} (\bibinfo{year}{2004}),
  \eprint{quant-ph/0408040}.

\bibitem[{\citenamefont{{Jelezko \it{et al.}}}(2004{\natexlab{a}})}]{jelezko04}
\bibinfo{author}{\bibfnamefont{F.}~\bibnamefont{{Jelezko \it{et al.}}}},
  \bibinfo{journal}{Phys. Rev. Lett.} \textbf{\bibinfo{volume}{92}},
  \bibinfo{pages}{076401} (\bibinfo{year}{2004}{\natexlab{a}}).

\bibitem[{\citenamefont{{Jelezko \it{et
  al.}}}(2004{\natexlab{b}})}]{jelezko04b}
\bibinfo{author}{\bibfnamefont{F.}~\bibnamefont{{Jelezko \it{et al.}}}},
  \bibinfo{journal}{Phys. Rev. Lett.} \textbf{\bibinfo{volume}{93}},
  \bibinfo{pages}{130501} (\bibinfo{year}{2004}{\natexlab{b}}).

\bibitem[{\citenamefont{Dur et~al.}(1999)\citenamefont{Dur, Briegel, Cirac, and
  Zoller}}]{Dur99}
\bibinfo{author}{\bibfnamefont{W.}~\bibnamefont{Dur}},
  \bibinfo{author}{\bibfnamefont{H.~J.} \bibnamefont{Briegel}},
  \bibinfo{author}{\bibfnamefont{J.~I.} \bibnamefont{Cirac}}, \bibnamefont{and}
  \bibinfo{author}{\bibfnamefont{P.}~\bibnamefont{Zoller}},
  \bibinfo{journal}{Phys. Rev. A.} \textbf{\bibinfo{volume}{59}},
  \bibinfo{pages}{169} (\bibinfo{year}{1999}).

\bibitem[{\citenamefont{Dur}(1998)}]{DurThesis}
\bibinfo{author}{\bibfnamefont{W.}~\bibnamefont{Dur}},
  \bibinfo{journal}{Masters Thesis}  (\bibinfo{year}{1998}).

\bibitem[{\citenamefont{{Gaebel \it{et al.}}}(2004)}]{Gaebel04}
\bibinfo{author}{\bibfnamefont{T.}~\bibnamefont{{Gaebel \it{et al.}}}},
  \bibinfo{journal}{New Journal of Physics} \textbf{\bibinfo{volume}{6}},
  \bibinfo{pages}{98} (\bibinfo{year}{2004}).

\bibitem[{Kil()}]{Kilin}
\bibinfo{note}{S. Kilin, private communication.}

\bibitem[{\citenamefont{{Kennedy \it{et al.}}}(2003)}]{Kennedy03}
\bibinfo{author}{\bibfnamefont{T.}~\bibnamefont{{Kennedy \it{et al.}}}},
  \bibinfo{journal}{Appl. Phys. Lett.} \textbf{\bibinfo{volume}{83}},
  \bibinfo{pages}{4190} (\bibinfo{year}{2003}).

\bibitem[{\citenamefont{{Imamoglu \it{et al.}}}(1999)}]{imamoglu99}
\bibinfo{author}{\bibfnamefont{A.}~\bibnamefont{{Imamoglu \it{et al.}}}},
  \bibinfo{journal}{Phys. Rev. Lett.} \textbf{\bibinfo{volume}{83}},
  \bibinfo{pages}{4204} (\bibinfo{year}{1999}).

\bibitem[{\citenamefont{Pazy et~al.}(2003)\citenamefont{Pazy, Biolatti,
  Calarco, D'Amico, Zanardi, Rossi, and Zoller}}]{pazy03}
\bibinfo{author}{\bibfnamefont{E.}~\bibnamefont{Pazy}},
  \bibinfo{author}{\bibfnamefont{E.}~\bibnamefont{Biolatti}},
  \bibinfo{author}{\bibfnamefont{T.}~\bibnamefont{Calarco}},
  \bibinfo{author}{\bibfnamefont{I.}~\bibnamefont{D'Amico}},
  \bibinfo{author}{\bibfnamefont{P.}~\bibnamefont{Zanardi}},
  \bibinfo{author}{\bibfnamefont{F.}~\bibnamefont{Rossi}}, \bibnamefont{and}
  \bibinfo{author}{\bibfnamefont{P.}~\bibnamefont{Zoller}},
  \bibinfo{journal}{Europhys. Lett.} \textbf{\bibinfo{volume}{62}},
  \bibinfo{pages}{175} (\bibinfo{year}{2003}), \eprint{cond-mat/0109337}.

\bibitem[{\citenamefont{{Bracker \it{et al.}}}(2004)}]{bracker04}
\bibinfo{author}{\bibfnamefont{A.~S.} \bibnamefont{{Bracker \it{et al.}}}},
  \bibinfo{journal}{e-print:cond-mat/0408466}  (\bibinfo{year}{2004}).

\bibitem[{\citenamefont{Golovach et~al.}(2003)\citenamefont{Golovach,
  Khaetskii, and Loss}}]{golovach03}
\bibinfo{author}{\bibfnamefont{V.}~\bibnamefont{Golovach}},
  \bibinfo{author}{\bibfnamefont{A.}~\bibnamefont{Khaetskii}},
  \bibnamefont{and} \bibinfo{author}{\bibfnamefont{D.}~\bibnamefont{Loss}},
  \bibinfo{journal}{eprint: cond-mat/0310655}  (\bibinfo{year}{2003}).

\bibitem[{\citenamefont{{Ramanathan \it{et al}}}(2004)}]{Ramanathan04}
\bibinfo{author}{\bibfnamefont{C.}~\bibnamefont{{Ramanathan \it{et al}}}},
  \bibinfo{journal}{eprint: quant-ph/0408166}  (\bibinfo{year}{2004}).

\bibitem[{\citenamefont{Taylor et~al.}(2003{\natexlab{a}})\citenamefont{Taylor,
  Marcus, and Lukin}}]{taylor03}
\bibinfo{author}{\bibfnamefont{J.~M.} \bibnamefont{Taylor}},
  \bibinfo{author}{\bibfnamefont{C.~M.} \bibnamefont{Marcus}},
  \bibnamefont{and} \bibinfo{author}{\bibfnamefont{M.~D.} \bibnamefont{Lukin}},
  \bibinfo{journal}{Phys. Rev. Lett.} \textbf{\bibinfo{volume}{90}},
  \bibinfo{pages}{206803} (\bibinfo{year}{2003}{\natexlab{a}}).

\bibitem[{\citenamefont{{Taylor \it{et al.}}}(2004)}]{taylor04}
\bibinfo{author}{\bibfnamefont{J.~M.} \bibnamefont{{Taylor \it{et al.}}}},
  \bibinfo{journal}{e-print: cond-mat/0407640}  (\bibinfo{year}{2004}).

\bibitem[{\citenamefont{{Kiraz \it{et al.}}}(2002)}]{kiraz02}
\bibinfo{author}{\bibfnamefont{A.}~\bibnamefont{{Kiraz \it{et al.}}}},
  \bibinfo{journal}{Phys. Rev. B} \textbf{\bibinfo{volume}{65}},
  \bibinfo{pages}{161303(R)} (\bibinfo{year}{2002}).

\bibitem[{\citenamefont{Taylor et~al.}(2003{\natexlab{b}})\citenamefont{Taylor,
  Imamoglu, and Lukin}}]{taylor03b}
\bibinfo{author}{\bibfnamefont{J.~M.} \bibnamefont{Taylor}},
  \bibinfo{author}{\bibfnamefont{A.}~\bibnamefont{Imamoglu}}, \bibnamefont{and}
  \bibinfo{author}{\bibfnamefont{M.}~\bibnamefont{Lukin}},
  \bibinfo{journal}{Phys. Rev. Lett.} \textbf{\bibinfo{volume}{91}},
  \bibinfo{pages}{246802} (\bibinfo{year}{2003}{\natexlab{b}}).

\bibitem[{\citenamefont{{Imamoglu \it{et al.}}}(2003)}]{imamoglu03}
\bibinfo{author}{\bibfnamefont{A.}~\bibnamefont{{Imamoglu \it{et al.}}}},
  \bibinfo{journal}{Phys. Rev. Lett.} \textbf{\bibinfo{volume}{91}},
  \bibinfo{pages}{017402} (\bibinfo{year}{2003}).

\end{thebibliography}

\end{document}